\documentclass{emulateapj}
\usepackage{epsfig}
\usepackage{graphicx}

\sfcode`A=1000 \sfcode`B=1000 \sfcode`C=1000 \sfcode`D=1000
\sfcode`E=1000 \sfcode`F=1000 \sfcode`G=1000 \sfcode`H=1000
\sfcode`I=1000 \sfcode`J=1000 \sfcode`K=1000 \sfcode`L=1000
\sfcode`M=1000 \sfcode`N=1000 \sfcode`O=1000 \sfcode`P=1000
\sfcode`Q=1000 \sfcode`R=1000 \sfcode`S=1000 \sfcode`T=1000
\sfcode`U=1000 \sfcode`V=1000 \sfcode`W=1000 \sfcode`X=1000
\sfcode`Y=1000 \sfcode`Z=1000

\newcommand{\hc}{$H_{160}$}

\newcommand{\SSS}{{\it Spitzer}}
\newcommand{\etal}{{et al.}~}

\newcommand{\Msun}{$M_\odot$}

\def\spose#1{\hbox to 0pt{#1\hss}}
\def\simlt{\mathrel{\spose{\lower 3pt\hbox{$\mathchar"218$}}
     \raise 2.0pt\hbox{$\mathchar"13C$}}}
\def\simgt{\mathrel{\spose{\lower 3pt\hbox{$\mathchar"218$}}
     \raise 2.0pt\hbox{$\mathchar"13E$}}}

\slugcomment{2015 May 22.  Accepted by ApJS.}

\shorttitle{Spitzer-CANDELS}

\begin{document}
\title{S-CANDELS: The Spitzer-Cosmic Assembly Near-Infrared Deep Extragalactic Survey.  Survey Design, Photometry, and Deep IRAC Source Counts}

\author
{
M.~L.~N.\,Ashby\altaffilmark{1},
S.~P.~Willner\altaffilmark{1}, 
G.~G.~Fazio\altaffilmark{1}, 
J.~S.~Dunlop\altaffilmark{2}, 
E.~Egami\altaffilmark{3}, 
S.~M.~Faber\altaffilmark{4}, 
H.~C.~Ferguson\altaffilmark{5}, 
N.~A.~Grogin\altaffilmark{5},
J.~L.~Hora\altaffilmark{1}, 
J.-S.~Huang\altaffilmark{1}, 
A.~M.~Koekemoer\altaffilmark{5}, 
I.~Labb\'e\altaffilmark{6}, 
and 
Z.~Wang\altaffilmark{1}
}

\altaffiltext{1}{Harvard-Smithsonian Center for Astrophysics, 60 Garden St., Cambridge, MA 02138, USA
[e-mail:  {\tt mashby@cfa.harvard.edu}]}
\altaffiltext{2}{Scottish Universities Physics Alliance, Institute for Astronomy, University of Edinburgh, Royal Observatory, Edinburgh, EH9 3HJ, UK}
\altaffiltext{3}{Steward Observatory, University of Arizona, 933 N. Cherry Ave, Tucson, AZ 85721, USA}
\altaffiltext{4}{University of California Observatories/Lick Observatory and Department of Astronomy and Astrophysics University of California Santa Cruz, 1156 High St., Santa Cruz, CA 95064, USA}
\altaffiltext{5}{Space Telescope Science Institute, 3700 San Martin Drive, Baltimore, MD 21218, USA}
\altaffiltext{6}{Leiden Observatory, Leiden University, NL-2300 RA Leiden, Netherlands}

\begin{abstract}

The \SSS-Cosmic Assembly Deep Near-Infrared Extragalactic Legacy Survey 
(S-CANDELS; PI G.\ Fazio) is a Cycle 8 Exploration Program designed to detect 
galaxies at very high redshifts ($z>5$).  To mitigate the effects of 
cosmic variance and also to take advantage of deep coextensive coverage 
in multiple bands by the {\sl Hubble Space Telescope} Multi-Cycle Treasury Program CANDELS, 
S-CANDELS was carried out within five widely separated extragalactic 
fields:  the UKIDSS Ultra-Deep Survey, 
the Extended {\sl Chandra} Deep Field South, COSMOS, the {\sl HST} Deep 
Field North, and the Extended Groth Strip.  
S-CANDELS builds upon the existing coverage of these fields from 
the Spitzer Extended Deep Survey (SEDS), a Cycle 6 Exploration Program,
by increasing the integration time from SEDS' 12 hours to a total
of 50 hours but within a smaller area, 0.16\,deg$^2$.  
The additional depth significantly
increases the survey completeness at faint magnitudes.
This paper describes the S-CANDELS survey design, processing,
and publicly-available data products.  
We present IRAC dual-band 3.6+4.5\,$\mu$m catalogs
reaching to a depth of 26.5\,AB mag.
Deep IRAC counts for
the roughly 135,000 galaxies detected by S-CANDELS are consistent with
models based on known galaxy populations.  The increase in 
depth beyond earlier \SSS/IRAC surveys does not reveal a significant
additional contribution from discrete sources to the 
diffuse Cosmic Infrared Background (CIB).  Thus it remains true that
only roughly half of the estimated CIB flux from COBE/DIRBE is
resolved.
\end{abstract}


\section{Introduction}

Deep imaging at infrared wavelengths is now a standard tool for detecting and
identifying galaxies at the highest redshifts (e.g., Oesch et al. 2013; 
Finkelstein et al. 2013).  Indeed, deep infrared surveys carried out in the
low-background conditions prevailing in space are indispensable for 
reliable detections of the most distant objects.  Moreover, observations
carried out in the infrared regime benefit from their sensitivity to
rest-frame stellar light, relatively free from attenuation by dust.  
Thus space-based infrared observations have a demonstrated capability to 
detect distant galaxies and characterize their stellar content.

The Infrared Array Camera (IRAC; Fazio et al.\ 2004) aboard the {\sl Spitzer
Space Telescope} (Werner et al.\ 2004) has made significant additions to
our knowledge of high-redshift galaxies.  Although {\sl Hubble Space Telescope} (HST)
observations have been essential for identifying candidate high-redshift galaxies
using the Lyman-break technique (e.g., Steidel et al.\ 1996a,b), infrared imaging by
IRAC, particularly in its 3.6 and 4.5\,$\mu$m bandpasses, has proved essential
for confirming the high-redshift nature of these objects and for understanding
the physical processes within them.  IRAC data enable
photometric redshift measurements and constrain stellar masses, ages, and
star formation histories.  IRAC has revealed, for example, that high-redshift
galaxies were suprisingly massive ($\sim10^{10}$\,\Msun) and had appreciable
stellar ages (200--300\,Myr), permitting new estimates of the
star formation rate in the early universe ($z=7-10$; e.g., Eyles et al.\ 2005; 
Egami et al.\ 2005; Yan et al.\ 2005; 2006; 2014; Labb\'e et al.\ 2006; 2007; 2010; 
2013; Stark et al. 2007).

\begin{figure}
\includegraphics[bb= 18 144 592 718,width=\columnwidth]{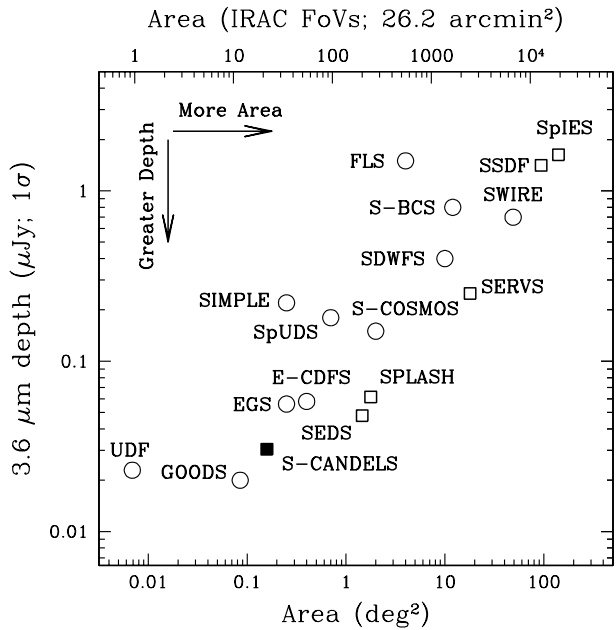}
\caption{
Comparison of measured S-CANDELS 3.6\,\micron\ 1$\sigma$ depth 
and total area (solid square) to other major \SSS/IRAC extragalactic surveys
from the cyrogenic mission (circles) and warm mission (squares). 
The points shown for S-CANDELS, SEDS (Ashby et al.\ 2013), 
SDWFS (Ashby et al.\ 2009), and SSDF (Ashby et al. 2013b) are based on 
photometry of simulated sources in mosaic images and therefore account 
for source confusion.  All other points are taken from the online calculator 
SENS-PET, using the appropriate sensitivies for the cryogenic and warm mission.  
Low-background conditions were assumed throughout except for SpIES, for which
a medium background was (conservatively) assumed.
Surveys shown include GOODS (Great Observatories Origins Deep Survey; Giavalisco et al.\ 2004; 
Wang et al.\ 2010; Hathi et al.\ 2012; Lin et al.\ 2012), 
the EGS (Extended Groth Strip; Davis et al.\ 2007; Bielby et al.\ 2012), 
E-CDFS (Extended Chandra Deep Field South; Rix et al.\ 2004; Castellano et al.\ 2010), 
SpUDS (\SSS\ Public Legacy Survey of UKIDSS Ultra-Deep Survey; Caputi et al. 2011),
SCOSMOS (\SSS\ Deep Survey of {\sl HST} COSMOS 2-Degree ACS Field; Scoville et al.\ 2007b),
SERVS (\SSS\ Extragalactic Representative Volume Survey; Mauduit et al.\ 2012),
BCS (Blanco Cluster Survey), 
SWIRE (\SSS\ Wide-area Infrared Extragalactic Survey; Lonsdale et al.\ 2003, 2004), 
the FLS (\SSS\ First Look Survey; Fang et al.\ 2004), 
the UDF (Ultra-Deep Field; Labb\'e et al.\ 2013), 
SIMPLE (the \SSS\ IRAC/MUSYC Public Legacy in E-CDFS; Damen et al.\ 2011), 
SpIES (\SSS-IRAC Equatorial Survey),
and SPLASH (\SSS\ Large-Area Survey with Hyper-Suprime-Cam).
Compared to the analogous Figure 1 of Ashby et al.\ (2013), 
SpIES, SSDF, SERVS, and S-CANDELS have all been updated because
they relied on SENS-PET estimates that did not yet
account for the slightly reduced sensitivity of IRAC during the warm mission.
}
\label{fig:etendue}
\end{figure}

The successes of deep surveys played a major role in motivating the {\sl HST} Multi-Cycle Treasury
Program known as the Cosmic Assembly Deep Near-Infrared Extragalactic Legacy Survey
(CANDELS), which used the Wide-Field Camera 3 (WFC3) to deeply cover five premier 
extragalactic survey fields both deeply and with high spatial resolution in the 
$YJH$ bands (Koekemoer et al.\ 2011; Grogin et al.\ 2011).  CANDELS also obtained 
roughly coextensive Advanced Camera for Surveys (ACS) parallel imaging at 
visible wavelengths.  All the CANDELS fields were also covered by IRAC 
at 3.6 and 4.5\,$\mu$m by the \SSS\ Extended Deep Survey (SEDS; 
Ashby et al.\ 2013), to furnish rest-frame visible-light detections of the most
distant objects detected by CANDELS.  Compared to CANDELS, SEDS covered a relatively
wide area (1.46\,deg$^2$ versus 0.16\,deg$^2$).  However, although SEDS is quite deep by
current survey standards (26\,AB mag, 3$\sigma$), it is not well-suited to
detect the faintest, most distant CANDELS sources.  For this reason our team 
has carried out a much deeper IRAC survey focused specifically on the CANDELS 
fields.  
The new observations were obtained as a 
{\it Spitzer} Cycle~8 Exploration Program called {\it Spitzer}-CANDELS 
(S-CANDELS; PI G.\ Fazio).  S-CANDELS achieved a total exposure time of
50\,hr in all CANDELS fields at both 3.6 and 4.5\,$\mu$m.  
Figure~\ref{fig:etendue} shows how S-CANDELS compares to the other major 
{\it Spitzer} surveys in terms of depth and coverage.

This paper documents and characterizes the S-CANDELS mosaics, which
are being publically released, and compares the results with SEDS. 
These objectives require using substantially the same methods
as for SEDS.  In particular, we use only the IRAC data for source
identification and photometry.  
Other groups within the CANDELS collaboration are using the
higher-resolution imaging from {\it HST}/WFC3 F160W as a prior for 
source identification (Sec.~\ref{sec:applications}). 
These efforts include
Galametz et al.\ (2013; for the UDS)
Guo et al.\ (2013; ECDFS),
Nayyeri et al.\ in preparation (COSMOS),
Barro et al.\ in preparation (HDFN), and
Stefanon et al.\ in preparation (EGS).  Of these, all but one
make use of the full-mission S-CANDELS mosaics created as
described below (Galametz et al.\ (2013) used the original SEDS data from
Ashby et al.\ 2013).
Our independence from other data sets also has the advantage 
of detecting extremely red sources that are invisible at shorter 
wavelengths, like those either thought to be at very high redshifts, or
to have extreme attenuation by dust.  Such sources exist and are being 
investigated (M.\ Stefanon et al.\ in preparation).

\begin{figure}
\includegraphics[bb=18 144 592 718,width=\columnwidth]{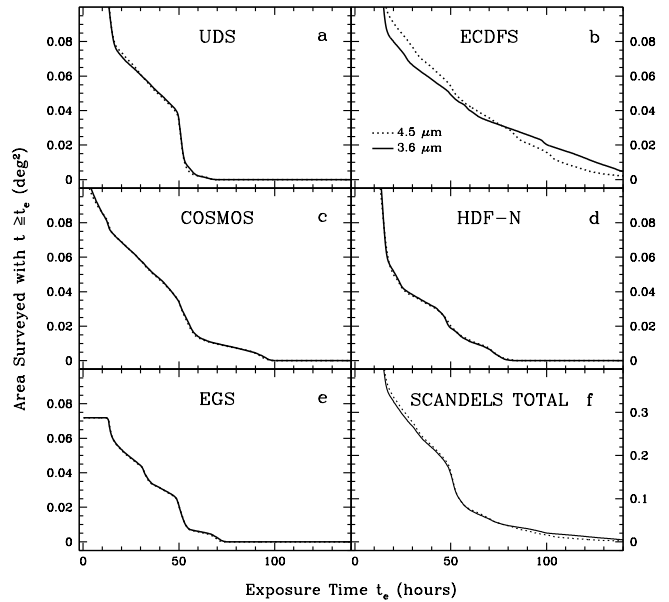}
\caption{
Cumulative area coverage as a function of exposure time for S-CANDELS, 
including other, earlier observations (Table~\ref{tab:obslog}).  
The solid and dotted lines correspond respectively to the 
3.6 and 4.5\,$\mu$m bands.  
Panel f shows the coverage summed over all five S-CANDELS fields.  
The nominal S-CANDELS depth was 50\,hr.
\label{fig:depths}}
\end{figure}

This paper is organized as follows.  Section~\ref{sec:obs} presents the observations;
Section~\ref{sec:fields} describes the individual S-CANDELS fields.  
Section~3 discusses the details of
the S-CANDELS 
source identification, photometry, and validation.  The results are
described in Section~\ref{sec:discussion}.
Section~\ref{sec:catalogs} describes the SEDS catalogs. 
Finally, Section~\ref{sec:applications} summarizes the benefits of the
50\,hr S-CANDELS depth and describes some uses of the data.
All magnitudes given in this paper are in the AB system.

\begin{deluxetable*}{rllr}
\tabletypesize{\footnotesize}
\tablewidth{344pt}
\tablecaption{The Five S-CANDELS Fields\label{tab:obslog}}
\tablehead{
{PID\tablenotemark{a}} & {Epoch} & {BCDs Used}  & Pipeline\\
                       &         & {3.6\,\micron,4.5\,\micron} & Version\\
}
\startdata
\multicolumn{4}{l}{UDS  (2:18:00, $-$5:10:17;  area=0.035,0.034\,deg$^2$)}\\
   181 & 2004 Jul 27--28          & 548,597\tablenotemark{b} & S18.7.0\\
 40021 & 2008 Jan 26--29          & 3640,3457 & " \\
 61041 & 2009 Sep 8--23           & 5255,5256 & S18.18.0 \\
 61041 & 2010 Feb 13--Mar 2       & 5328,5328 & " \\
 61041 & 2010 Sep 22--Oct 13      & 5436,5436 & " \\
 80218 & 2012 Feb 29--Mar 11      & 4680,4680 & S19.1.0 \\
 80218 & 2012 Oct 11--Oct 29      & 5328,5328 & " \\
 80218 & 2013 Mar 16              &  648, 648 & " \\

\multicolumn{4}{l}{ECDFS  (3:32:20, $-$27:37:20; area=0.049,0.054\,deg$^2$)}\\
    81 & 2004 Feb 16              & 167,146   & S18.7.0 \\
   194 & 2004 Feb 8--16           & 1724,1723\tablenotemark{c} & " \\
   194 & 2004 Aug 12--18          & 1632,1632\tablenotemark{c} & " \\
 20708 & 2005 Aug 19--23          & 1943,1872 & " \\
 20708 & 2006 Feb 6--11           & 1899,1944 & " \\
 30866 & 2007 Feb 15              & 1200,1080 & " \\
 60022 & 2010 Sep 20--Oct 4       & 4752,4588 & S18.18.0 \\
 70145 & 2010 Sep 16--Oct 25      & 3510,3510 & " \\
 70145 & 2011 Feb 11--Apr 7       & 4140,4140 & " \\
 70145 & 2011 Sep 21--Sep 22      &  630,630  & " \\
 70204 & 2011 Mar 17--Apr 7       & 5184,5128 & " \\
 60022 & 2011 Mar 26--Apr 7       & 4596,4752 & " \\
 80217 & 2011 Sep 25--Sep 28      & 1944,1944 & S19.1.0  \\
 60022 & 2011 Oct 10--Oct 20      & 4717,4552 & " \\
 80217 & 2012 Mar 30--Apr 5       & 1944,1943 & " \\

\multicolumn{4}{l}{COSMOS (10:00:30, +2:10:00;  area=0.034,0.034\,deg$^2$) }\\
 20070 & 2005 Dec 30--2006 Jan 2  & 1259,1253 & S18.7.0  \\
 61043 & 2010 Jan 25--Feb 4       & 3672,3672 & S18.18.0 \\
 61043 & 2010 Jun 10--28          & 3164,3140 & " \\
 61043 & 2011 Jan 30--Feb 6       & 3180,3196 & " \\
 80057 & 2012 Feb  4--Feb 19      & 6840,6840 & S19.1.0  \\
 80057 & 2012 Jun 26--Jul 9       & 6840,6840 & " \\

\multicolumn{4}{l}{{HDFN (12:36:12, +62:14:12; area=0.019,0.020\,deg$^2$)} }\\
    81 & 2004 May 26--27          & 215,178   & S18.7.0  \\
   169 & 2004 May 16--26          & 2609,2609\tablenotemark{c} & " \\
   169 & 2004 Nov 17--25          & 2447,2447\tablenotemark{c} & " \\
   169 & 2005 Nov 25              &  114,114\tablenotemark{c}  & " \\
 20218 & 2005 Nov 28--Dec 9       & 200,200   & " \\
 20218 & 2006 Jun 2--3            & 200,200   & " \\
 61040 & 2010 May 12-29           & 4895,4896 & S18.18.0 \\
 61040 & 2011 Feb 28--Mar 13      & 5440,5440 & " \\
 60140 & 2011 May 22--Jun 2       & 5208,4896 & " \\
 80215 & 2012 Jan 25--28          & 1872,1872 & S19.1.0  \\
 80215 & 2012 Jul 23--30          & 1944,1944 & " \\

\multicolumn{4}{l}{EGS (14:19:38, +52:25:47; area=0.021,0.021\,deg$^2$)}\\
     8 & 2003 Dec 21--28          & 988,969\tablenotemark{c}  & S18.7.0  \\
     8 & 2004 Jun 28--Jul 3       & 1027,989\tablenotemark{c} & "  \\
     8 & 2006 Mar 28--29          & 117,24\tablenotemark{c}   & " \\
 41023 & 2008 Jan 24--25          & 726,726   & " \\
 41023 & 2008 Jul 21--23          & 726,726   & " \\
 61042 & 2010 Feb 5--16           & 4056,4056 & S18.18.0  \\
 61042 & 2010 Aug 4--19           & 4021,4056 & " \\
 61042 & 2011 Feb 10--22          & 3970,4048 & " \\
 80216 & 2011 Aug 18--21          & 2052,2052 & S19.1.0   \\
 80216 & 2012 Feb  2--26          & 3888,3888 & " \\
 80216 & 2012 Aug 28--31          & 1836,1836 & " \\
\enddata
\tablecomments{S-CANDELS field positions and areas.  Areas given were covered 
respectively at 3.6 and 4.5\,$\mu$m to a depth of at least 50\,hours total 
integration time by the combined
sum of all programs listed here.  Compare to Table~\ref{fig:depths}.}
\tablenotetext{a}{\SSS\ Program Identification Number}
\tablenotetext{b}{30\,s frames.}
\tablenotetext{c}{200\,s frames.}
\end{deluxetable*}

\section{Observations and Data Reduction}
\label{sec:obs}

\begin{figure*}
\includegraphics[bb=75 257 550 535, width=\textwidth]{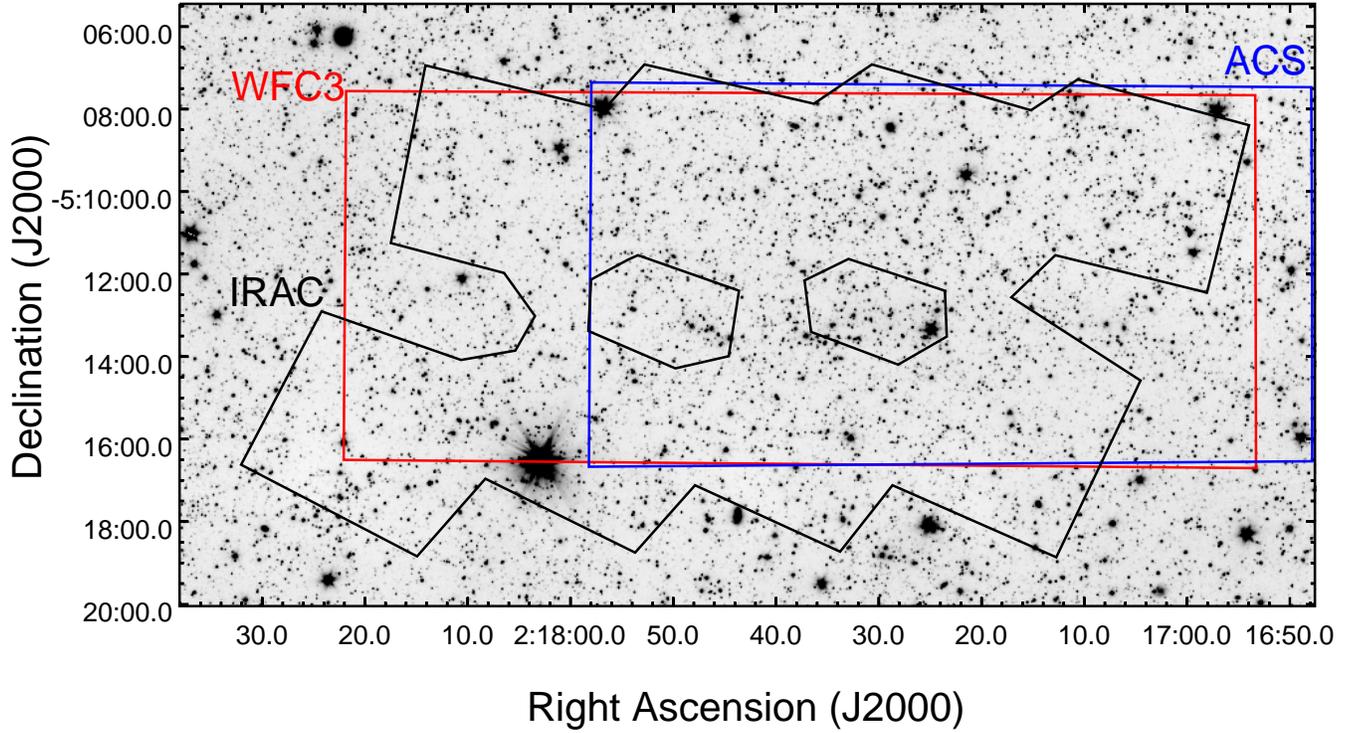}
\caption{The full-depth S-CANDELS 3.6\,$\mu$m mosaic in the UDS field, including exposures
from the cryogenic mission and SEDS.  
The image stretch ranges from -0.01 (white) to 0.05\,MJy/sr (black).
The minimum total exposure time in the field 
shown is 12\,hours.  The black polygons indicate where the total IRAC 
integration reaches 50\,hours.  The six-sided black polygons in the field center
enclose regions where the coverage grades down to the 12\,hour SEDS depth.
The red and blue rectangles respectively 
indicate the extent of the {\sl HST}/WFC3 and ACS imaging from CANDELS 
(Grogin et al.\ 2011; Koekemoer et al. 2011).  
\label{fig:UDS1}}
\end{figure*}

\begin{figure*}
\epsscale{1.05}
\includegraphics[bb=75 257 550 535, width=\textwidth]{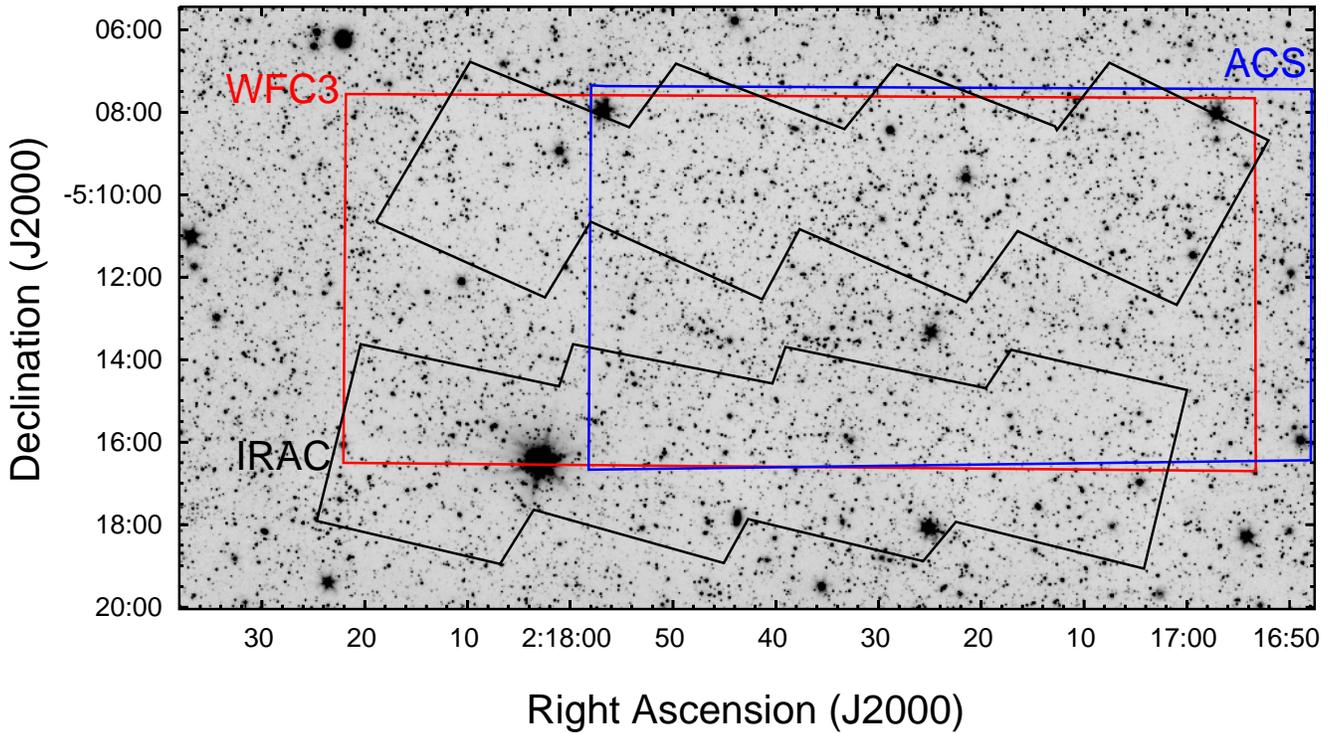}
\caption{
As Figure~\ref{fig:UDS1}, but showing the S-CANDELS 4.5\,$\mu$m mosaic of the UDS.
The stretch ranges from -0.01 to 0.05\,MJy\,sr$^{-1}$.
\label{fig:UDS2}}
\end{figure*}

\begin{figure}
\epsscale{1.15}
\includegraphics[bb=80 220 440 560, width=\columnwidth]{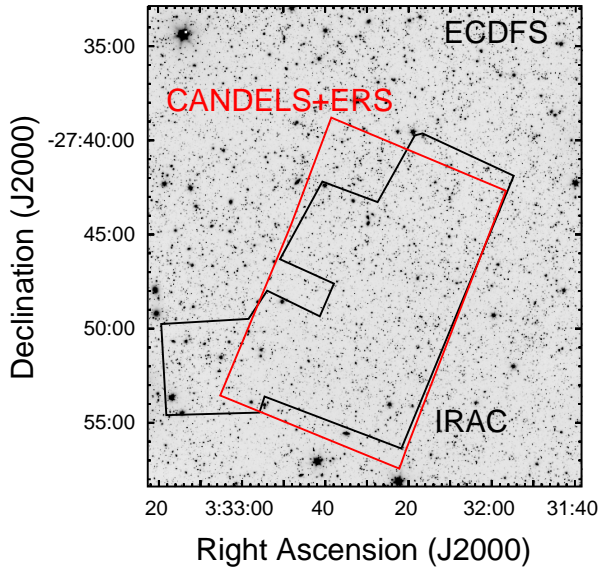}
\caption{The full-depth SCANDELS 3.6\,\micron\ mosaic of the ECDFS.  
The stretch ranges from -0.01 to 0.05\,MJy\,sr$^{-1}$.  The
black polygon encloses a region where the depth of coverage is a minimum of 
50\,hours.  The red rectangle indicates the portion of the field covered by
the {\sl HST} CANDELS and ERS programs.
\label{fig:ECDFS1}}
\end{figure}

\begin{figure}
\epsscale{1.15}
\includegraphics[bb=80 220 440 560, width=\columnwidth]{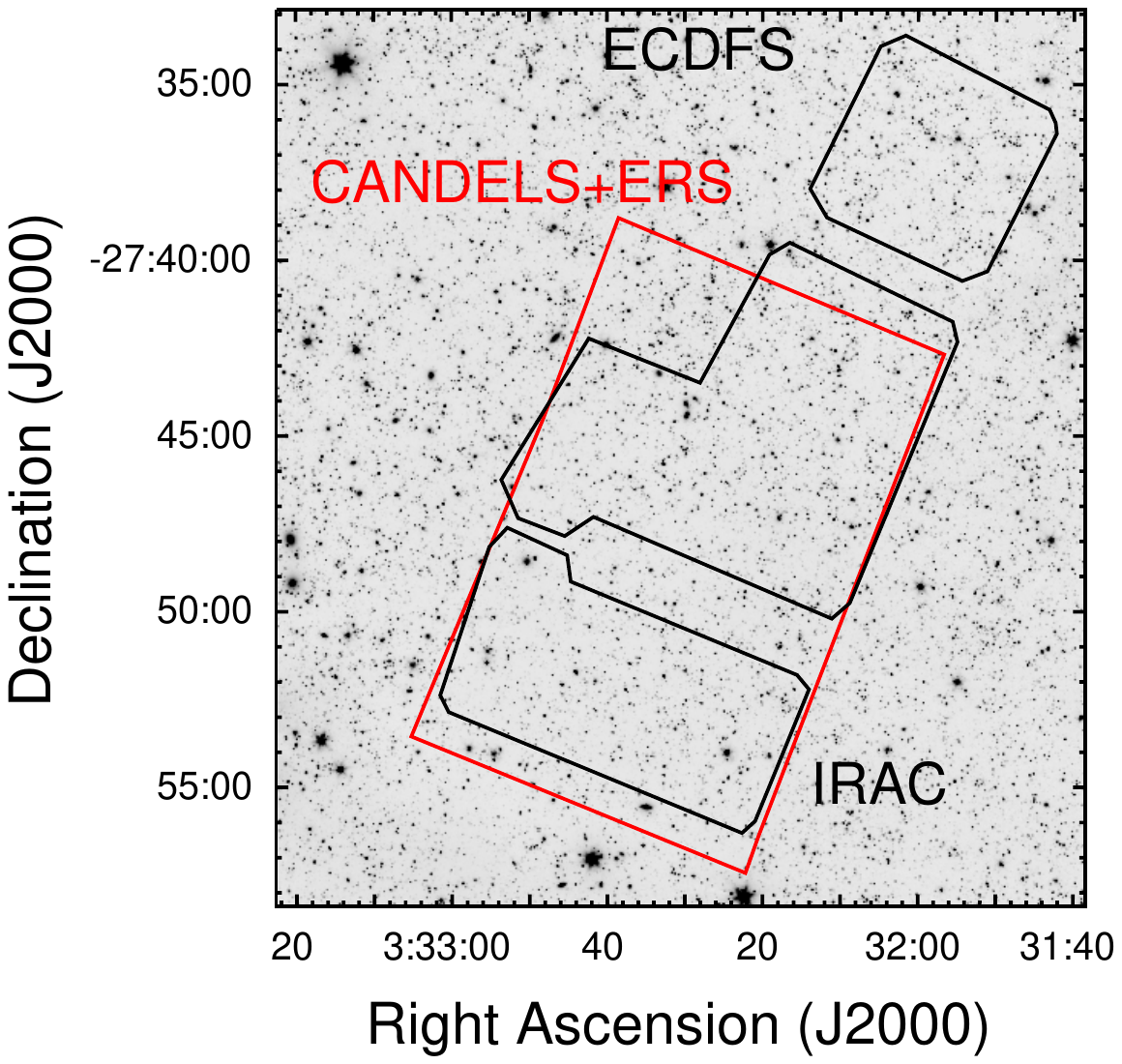}
\caption{As Figure~\ref{fig:ECDFS1}, but showing the full-depth 4.5\,\micron\ mosaic of 
the ECDFS including all SEDS and cryogenic imaging by IRAC.
The stretch ranges from -0.01 to 0.05\,MJy\,sr$^{-1}$.
\label{fig:ECDFS2}}
\end{figure}

\subsection{The Five S-CANDELS Survey Fields}
\label{sec:fields}
Because the scientific emphasis of S-CANDELS is on detecting and characterizing
galaxies at very high redshifts, it is vital that the S-CANDELS fields be
placed where sensitive photometry is available in multiple bands other than the
IRAC 3.6 and 4.5\,$\mu$m filters.  NIR and visible imaging deep enough to match
the IRAC observations reported here are of special importance.  
Accordingly, we chose to locate S-CANDELS
inside the wider and shallower fields already covered by SEDS, in regions
that enjoy deep optical and NIR imaging from {\sl HST}/CANDELS. 
These S-CANDELS fields are thus the Extended GOODS-South 
(aka the GEMS field, hereafter ECDFS; Rix \etal 2004; Castellano \etal 2010), 
the Extended GOODS-North (HDFN; Giavalisco \etal 2004; 
Wang \etal 2010; Hathi \etal 2012, Lin \etal 2012),
the UKIDSS Ultra-Deep Survey (UDS; aka the Subaru/XMM
Deep Field, Ouchi \etal 2001; Lawrence \etal 2007),
a narrow field within the Extended Groth Strip (EGS; Davis \etal 2007; 
Bielby \etal 2012), and a strip within the UltraVista deep survey
of the larger COSMOS field (Scoville \etal 2007a; McCracken \etal 2012).
These five S-CANDELS fields are distributed in ecliptic longitude and declination to
permit ground-based followup from both hemispheres.  

\begin{figure*}
\epsscale{0.8}
\includegraphics[bb=145 240 405 560,width=\textwidth]{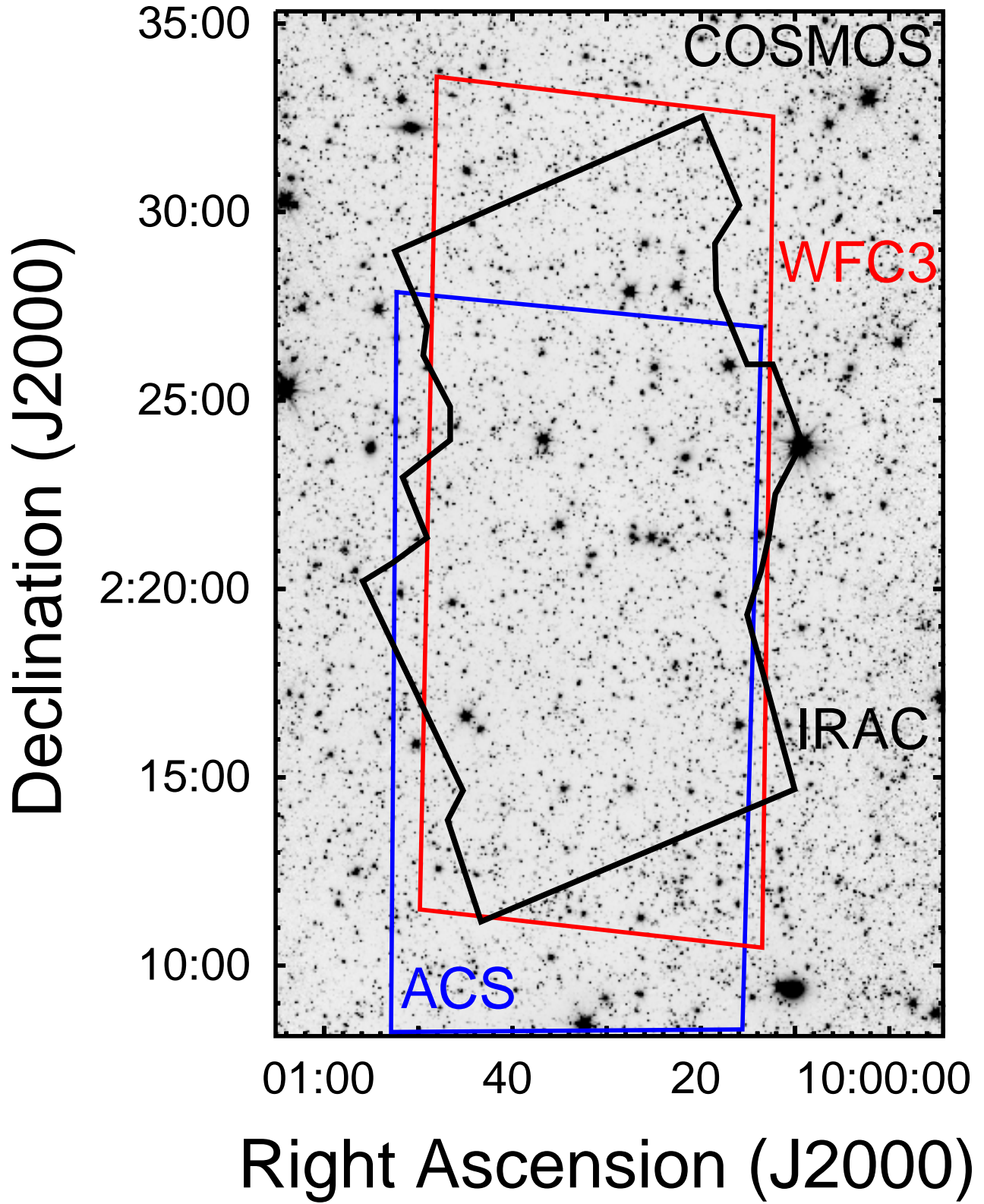} 
\caption{The total S-CANDELS 3.6\,$\mu$m mosaic in the COSMOS field including
all observations from SEDS and the cryogenic mission (Table~\ref{tab:obslog}).
The image stretch ranges from -0.01 (white) to 0.05\,MJy/sr (black).  The black
polygon approximately indicates the area covered by at least 50\,hours of 
integration time.  All of the field shown is covered by at least 12\,hours
total integration time.  The red line encloses the region covered by the CANDELS
WFC3 observations, and the blue line encloses the region observed by the 
CANDELS parallel ACS exposures.
\label{fig:COSMOS1}}
\end{figure*}

\begin{figure*}
\epsscale{0.8}
\includegraphics[bb=145 240 405 560,width=\textwidth]{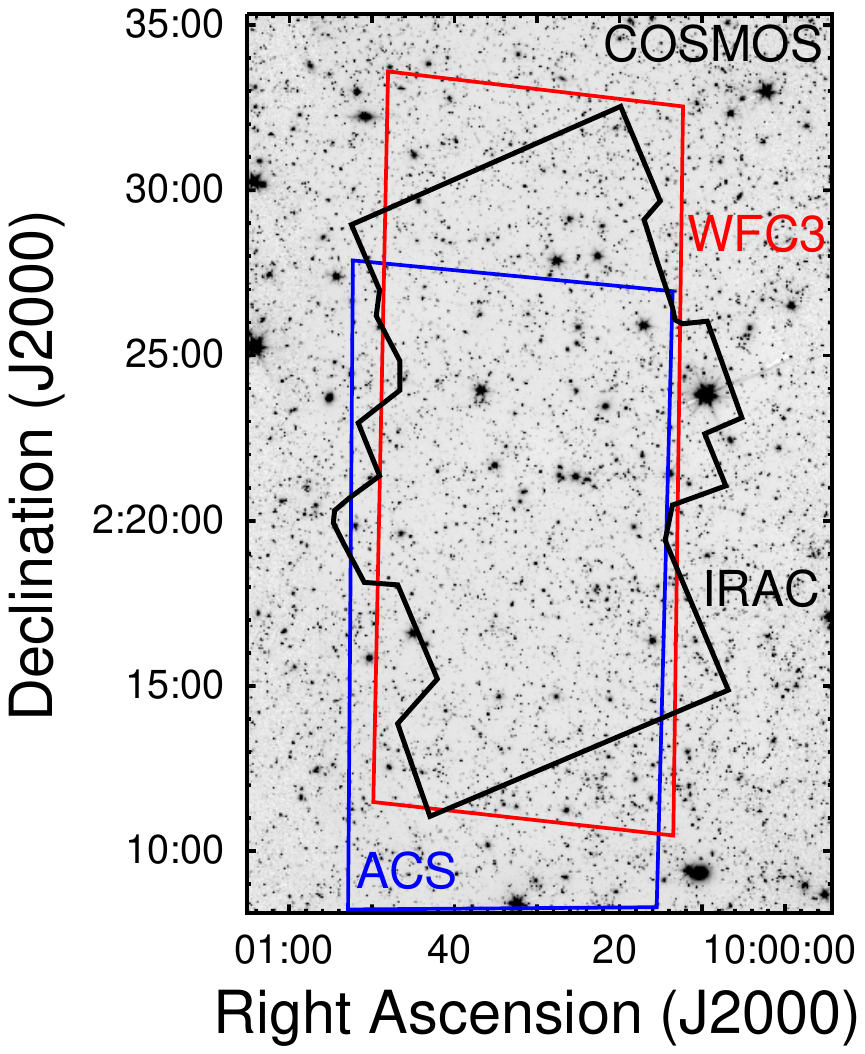}
\caption{As Figure~\ref{fig:COSMOS1}, but showing the full-depth 4.5\,$\mu$m mosaic 
in the COSMOS field.  The stretch ranges from -0.01 to 0.05\,MJy/sr.
\label{fig:COSMOS2}}
\end{figure*}

\subsection{Mapping Strategy}

The depths, areas, and sensitivity of earlier IRAC coverage of the five 
S-CANDELS fields up to and including the SEDS campaigns are described by Ashby 
et al.\ (2013).  The S-CANDELS observations were of a similar character, but had
a different etendue.  Figure~\ref{fig:depths} shows the cumulative depth vs area plots
for S-CANDELS, which had a design depth of 50\,hr.   

\begin{figure}
\epsscale{1.2}
\includegraphics[bb=85 235 440 560,width=\columnwidth]{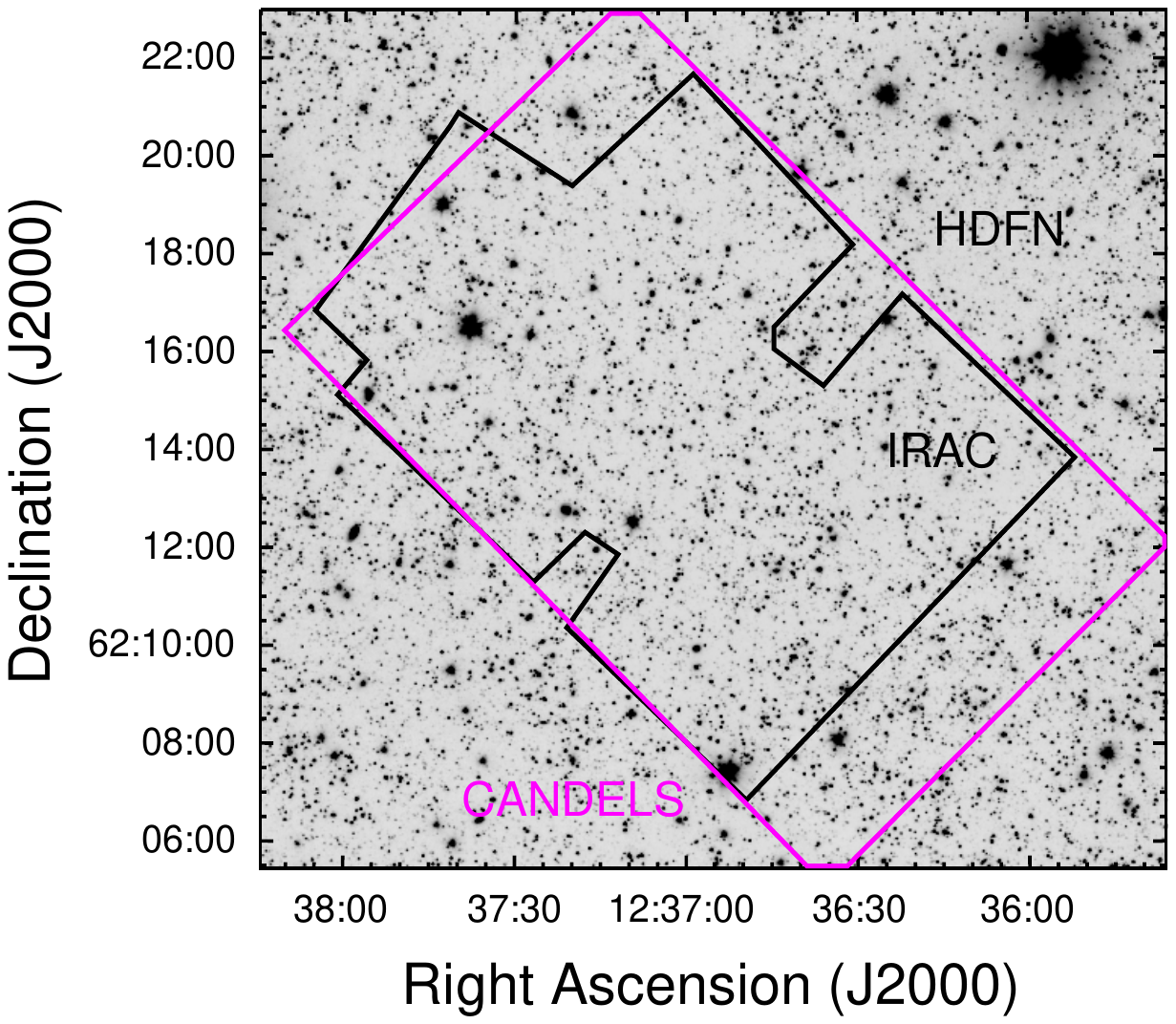}
\caption{Total SCANDELS IRAC 3.6\,\micron\ mosaic in the HDFN. 
The black polygon indicates approximately where the total 3.6\,\micron\
integration time, which includes observations from the cryogenic mission 
(i.e., GOODS), rises to at least 50\,hours.  The magenta rectangle 
indicates the coextensive {\sl HST}/WFC3+ACS footprint from CANDELS.
The stretch ranges from -0.01 to 0.05\,MJy\,sr$^{-1}$.
\label{fig:HDFN1}}
\end{figure}

\begin{figure}
\epsscale{1.2}
\includegraphics[bb=85 235 440 560,width=\columnwidth]{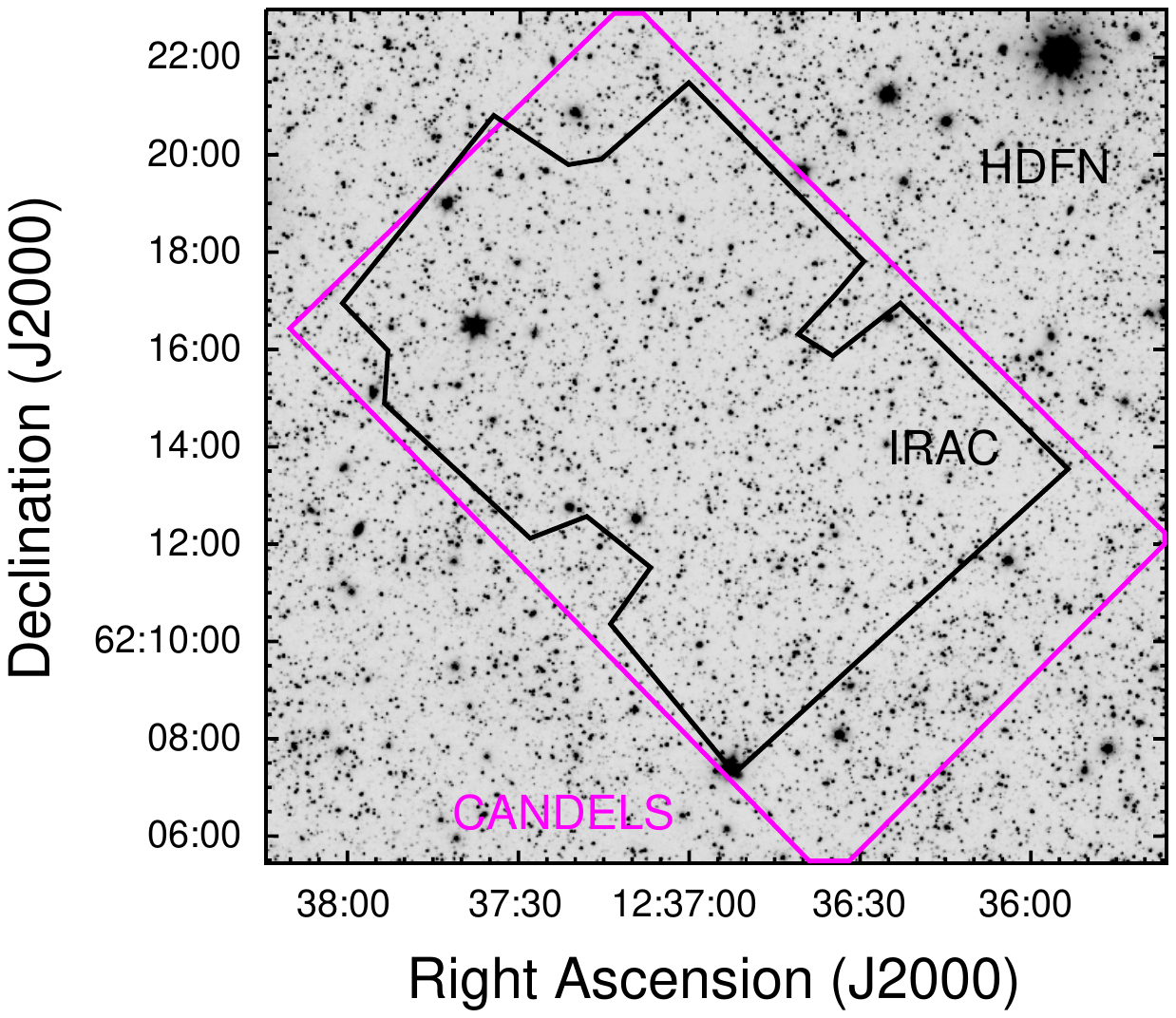}
\caption{As Figure~\ref{fig:HDFN1}, but for the 4.5\,$\mu$m observations.
The stretch ranges from -0.01 to 0.05\,MJy\,sr$^{-1}$.
\label{fig:HDFN2}}
\end{figure}

The S-CANDELS observing strategy was designed to maximize the area
covered to full depth within the CANDELS \hc\ area.  Each field was
visited twice\footnote{Three AORs in the UDS field observed in 2012
  March were useless because solar particles saturated the
  detectors. These AORs were reobserved in 2013 March.  Recovery from
a spacecraft anomaly in 2011 August prevented observation of 17 AORs in the
  EGS. They were observed in 2012 August.} with six months separating
the two visits.
Table~\ref{tab:obslog} lists the epochs for each field.  All of the IRAC full-depth
coverage is within the SEDS area (Ashby et al.\ 2013), and almost all is
within the area covered by {\sl HST} for CANDELS.  (See Figs.~\ref{fig:UDS1}--\ref{fig:EGS2}.)

Each of the two observation epochs accumulated 19\,hr integration
time per pointing, or less when a field had pre-existing coverage
other than SEDS.  For efficiency, each position in the field was
usually observed for 2 frames of 100\,s each before moving the
telescope to the next position.  The medium Reuleaux 36 dither
pattern was used throughout, except for the EGS field, 
which used an 18-point dither
pattern.\footnote{The EGS dither pattern was equivalent to alternate
  positions of the medium Reuleaux 36 pattern and was specified via a
  cluster target.} The AORs thus sampled each sky position at 
many positions on the arrays. Each AOR (pair of linked AORs for
the EGS) covered the full intended field of view (FOV) {\em for one
  wavelength}, but the 3.6 and 4.5\,$\mu$m coverage did not overlap
for UDS, HDFN, or CDFS. For COSMOS and EGS, the overlap was only
partial.  However, the IRAC fields of view switch places every six
months, so the area observed at 3.6\,$\mu$m in one epoch was
observed at 4.5\,$\mu$m in the alternate epoch and vice versa to
achieve complete coverage of the intended area at both wavelengths.

\subsection{Data Reduction}

We used the same procedures to reduce the S-CANDELS data as were
applied earlier to the SEDS observations described by Ashby et al.\ (2013).
In the following we therefore describe only the most important 
aspects of the S-CANDELS reductions.  

All suitable
data were combined into full-mission mosaics that include coextensive 
imaging from SEDS and other projects from both the cryogenic and
warm missions (see Table~\ref{tab:obslog} for the complete lists) 
into full-mission mosaics covering the CANDELS fields.  
Processing was based on IRAC Corrected Basic Calibrated Data (cBCD) 
exposures generated by the pipeline versions indicated in 
Table~\ref{tab:obslog}.  The different pipeline versions differ
only in matters of minor artifact correction, not in overall 
calibration\footnote{\small http://irsa.ipac.caltech.edu/docs/irac/iracinstrument
handbook/73/} 
of the 3.6 or 4.5\,$\mu$m exposures.

\begin{figure*}
\includegraphics[bb=110 200 530 610,width=\textwidth]{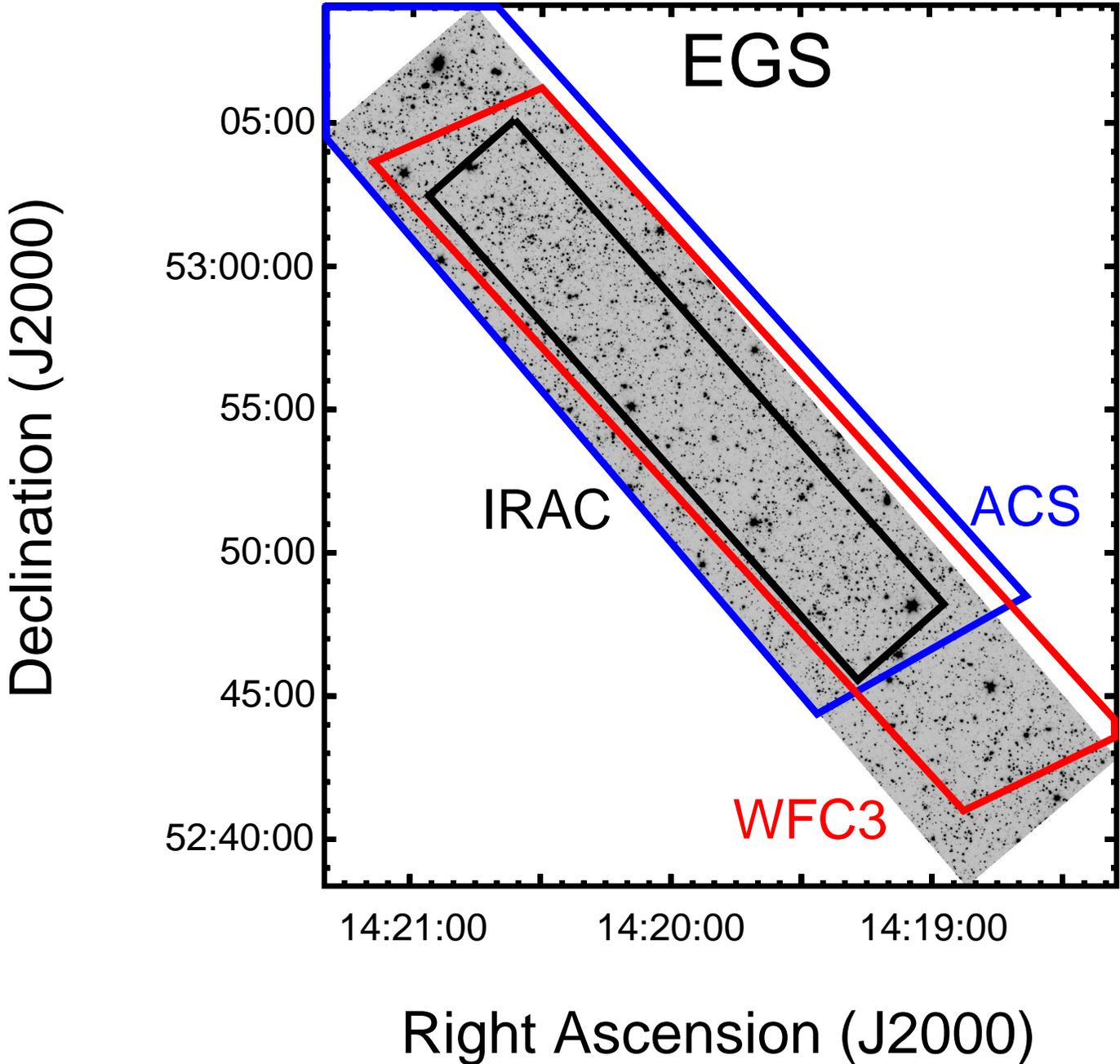}
\caption{Total S-CANDELS 3.6\,\micron\ mosaic of the EGS field.
The deepest coverage, at least 50\,hours total integration time, lies
within the black rectangle.
The stretch ranges from -0.01 to 0.05\,MJy\,sr$^{-1}$.
Outside the black rectangle the 3.6\,\micron\ integration time is at 
least 12\,hour.
The blue and red polygons respectively indicate the approximate locations of the 
CANDELS ACS and WFC3 coverage.
\label{fig:EGS1}}
\end{figure*}

\begin{figure*}
\epsscale{0.9}
\includegraphics[bb=100 240 430 560,width=\textwidth]{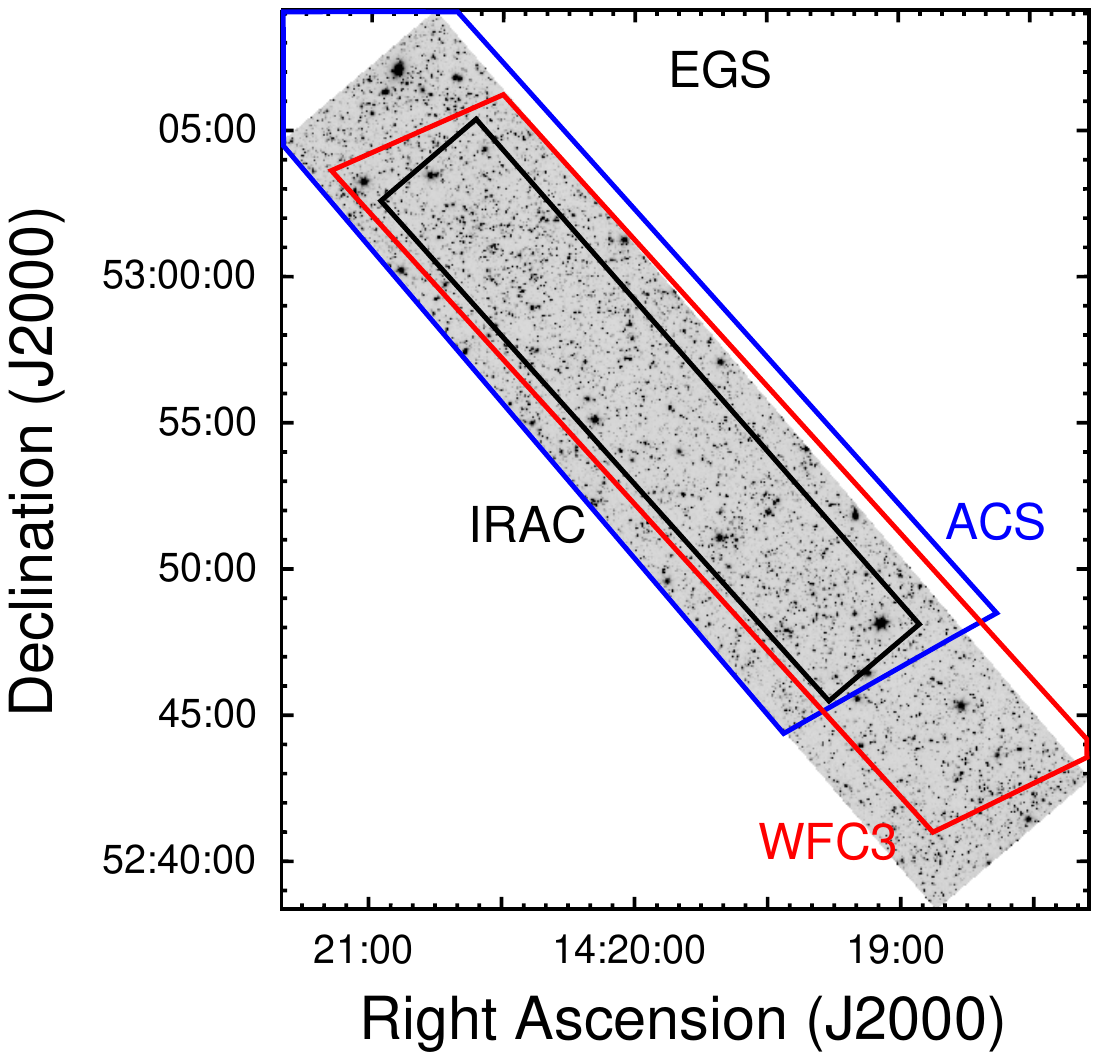}
\caption{As Figure~\ref{fig:EGS1}, but for the total 
S-CANDELS 4.5\,\micron\ mosaic of the EGS field.
The stretch ranges from -0.01 to 0.05\,MJy\,sr$^{-1}$.
\label{fig:EGS2}}
\end{figure*}

Before mosaicking, all the IRAC 
exposures were corrected for long-term residual images and for column pulldown.
The mosaics were constructed
with {\tt IRACproc} (Schuster \etal 2006) in the same way as was done
for SEDS, but over narrower fields.  In the ECDFS and the HDFN, which have
very large datasets, computer memory constraints prevented us from making the
mosaics in a single {\tt IRACproc} run.  For these fields we mosaicked 
subsets of the exposures and subsequently mean-averaged the results into
a single mosaic.  As with SEDS, all the S-CANDELS mosaics were pixellated
to 0\farcs6 and were aligned to the tangent-plane projections used 
by the CANDELS team (Table\,5 of Koekemoer et al.\ 2011).  Figures~\ref{fig:UDS1} 
through \ref{fig:EGS2} show the final IRAC mosaics for all five fields.

The final mosaics, coverage maps, model images, and residual images 
are all available from the \SSS\ Exploration Science 
Programs website.\footnote{\small http://irsa.ipac.caltech.edu/data/SPITZER/docs/spitzer
mission/observingprograms/es/}  

\section{Source Extraction and Photometry}
\subsection{Source Identification}
\label{ssec:extraction}

Source confusion is pronounced even for the 12-hour SEDS mosaics; the problem 
impacts the deeper S-CANDELS mosaics even more strongly.  
We therefore used {\tt StarFinder} (ver.\ 1.6f; Diolaiti \etal 2000) 
to identify sources because {\tt StarFinder} 
is optimized for identification and photometry of heavily blended sources 
in crowded fields (e.g., globular cluster stars).  As was done for SEDS, 
the S-CANDELS catalogs were constructed in two steps.  First, 
{\tt StarFinder} was used to identify and locate sources (even faint, blended ones).  
Second, a custom code was used to correct biases in the {\tt StarFinder} photometry.

\begin{figure*}
\includegraphics[bb=36 234 577 558,width=\columnwidth]{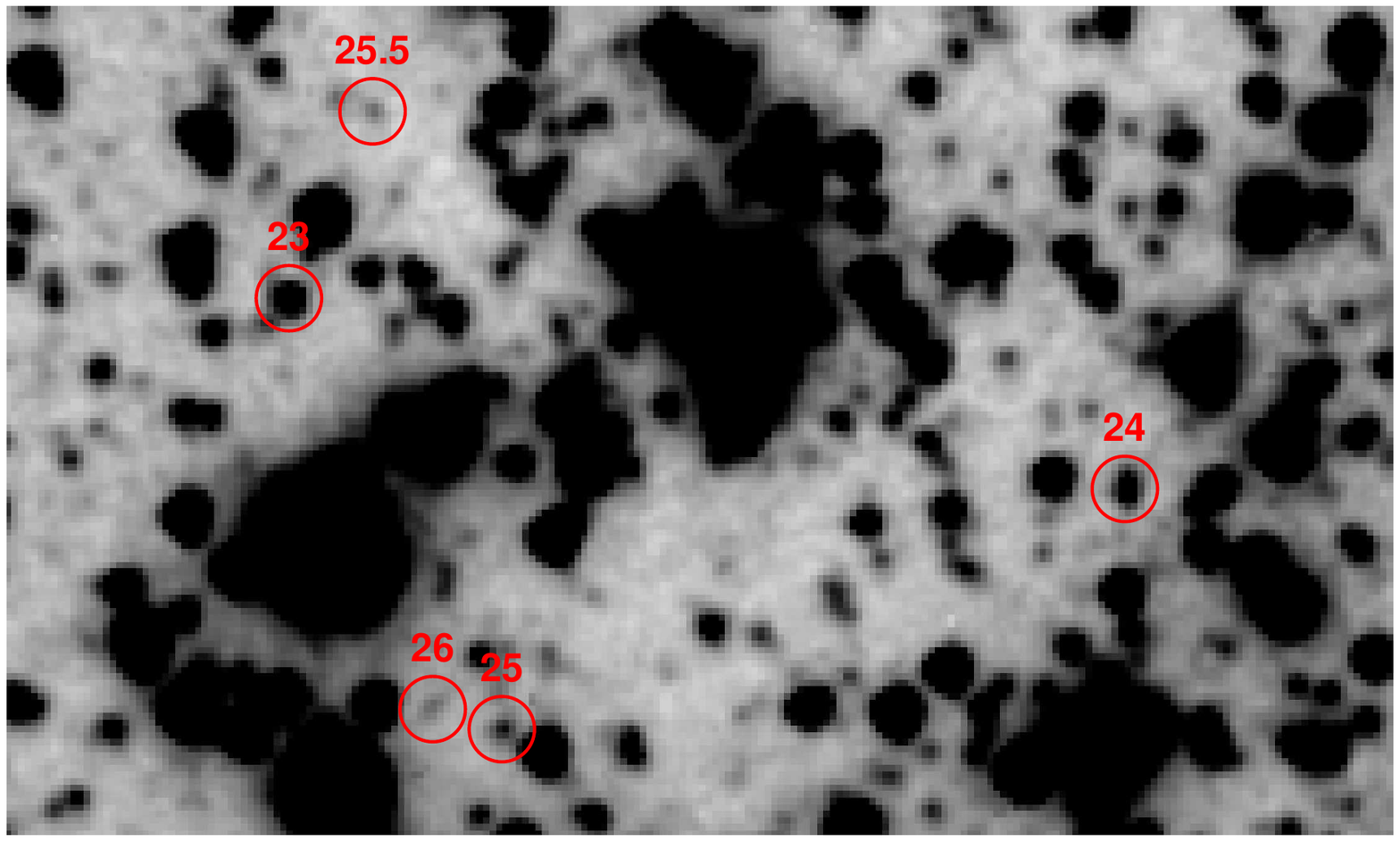}
\includegraphics[bb=36 234 577 558,width=\columnwidth]{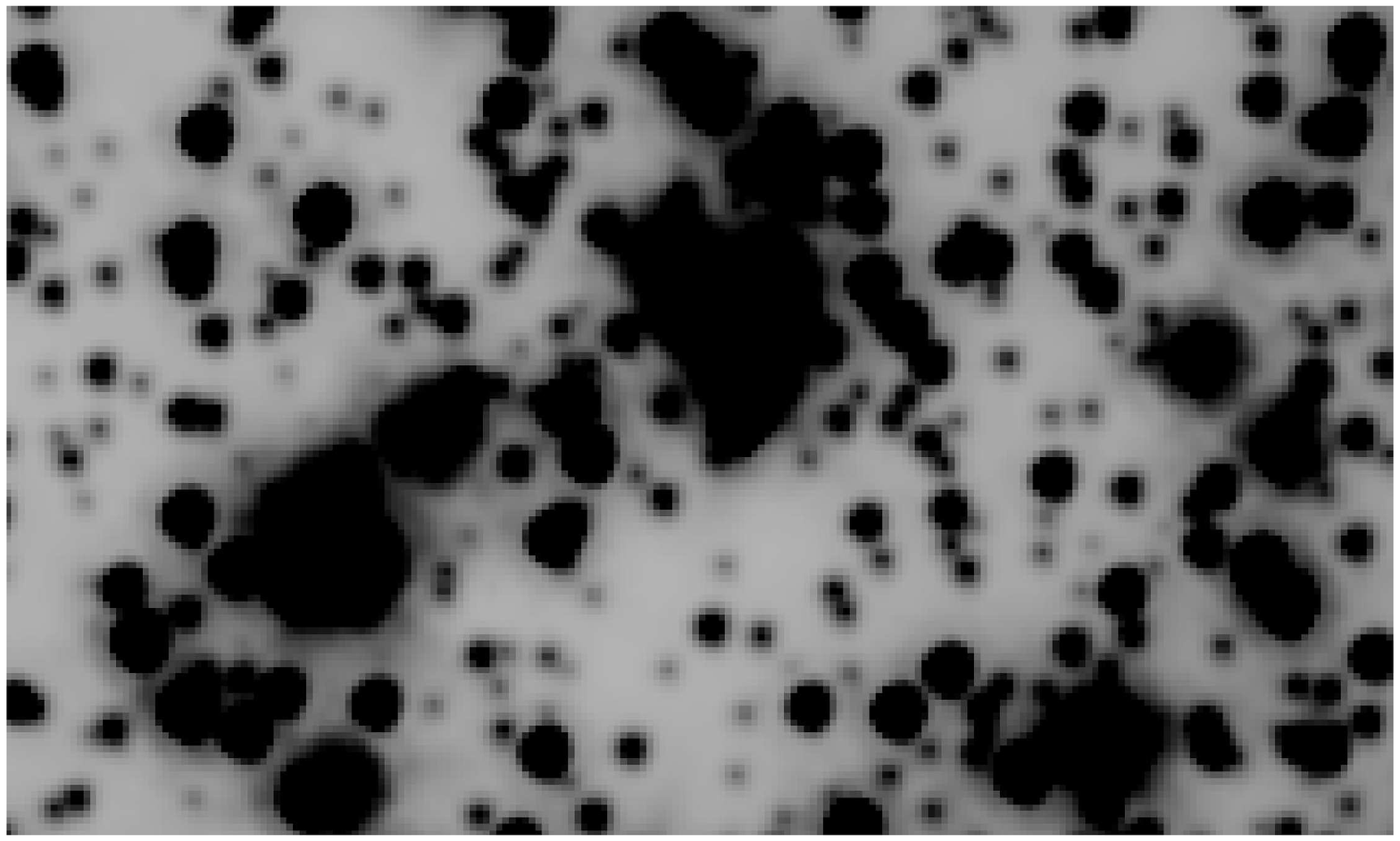}
\includegraphics[bb=36 234 577 558,width=\columnwidth]{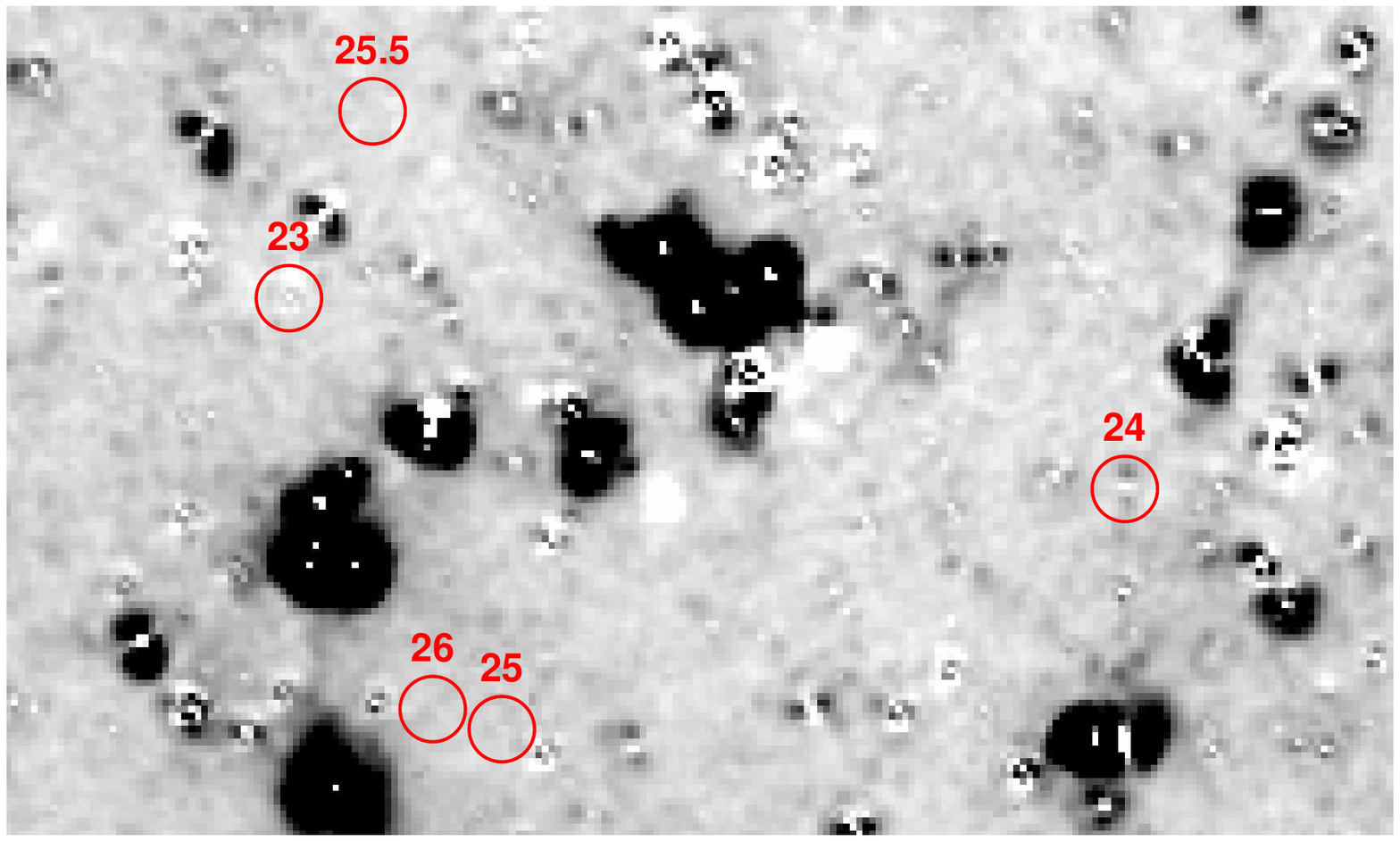}
\caption{Illustration of {\tt StarFinder} source extraction.
These negative images show a $2\farcm1\times1\farcm2$ region of the 
3.6\,$\mu$m S-CANDELS ECDFS mosaic; the other S-CANDELS band and 
fields behave in a similar way.  
The linear stretch ranges from $-$0.001 (white) to 0.004\,MJy\,sr$^{-1}$ (black)
throughout.
{\sl Top left:} The IRAC 3.6\,$\mu$m mosaic created by {\tt IRACProc}.
Red circles are placed around sources with AB magnitudes as shown.
{\sl Top Right:} The corresponding {\tt StarFinder} model image.
{\sl Bottom:} The residual image obtained after
the detected sources are removed.
\label{fig:resid}
}
\end{figure*}

The source-identification step was performed on the full-depth S-CANDELS mosaics.  
{\tt StarFinder} implements an algorithm based on iterated fitting and subtracting 
of a template point spread function (PSF) image.  Because the 
S-CANDELS mosaics are small and heavily confused, we were unable to identify 
enough isolated, sufficiently bright point sources to construct useful PSFs 
template images from the S-CANDELS mosaics themselves.  Instead, we used 
the PSF template images constructed earlier for SEDS.  Because the fields were
observed at similar roll angles for both SEDS and S-CANDELS, this should not
introduce significant error.  
{\tt StarFinder} is capable of repeating the source-identification algorithm
using the residuals it generates from a first pass through the mosaics.  This
allows the code to refine (reduce) its estimate of the background noise in the 
absence of the brightest objects.  
We therefore configured {\tt StarFinder} to process each field three times,
as was done for SEDS, with identical parameter settings.  In particular, the 
software was not allowed to deblend sources closer together than 0\farcs9, 
roughly half the FWHM of the PSF in the two warm IRAC passbands.
Based on our inspections of the final (third-pass) residual images 
(Figure~\ref{fig:resid} shows an example), 
we judged this approach successful.  The residual images are manifestly free
of large-scale background artifacts, and the faint sources (which are all 
effectively point sources) are well-fitted by our approach.

As was done for SEDS, the aperture magnitudes 
for each source were measured after re-inserting its fitted PSF 
at its fitted position in the {\tt StarFinder} residual image and then
measuring background-subtracted fluxes interior to 
diameters of 2\farcs4, 3\farcs6, 4\farcs8, 6\farcs0,
7\farcs2, and 12\farcs0.  Thus the S-CANDELS aperture magnitudes are relatively 
free of contamination by nearby neighbors (because to a good approximation they 
have been subtracted off) of both the photometered source and the nearby background.
The S-CANDELS catalogs contain both the PSF-fitted magnitudes 
based on the iterative procedure described above and the aperture magnitudes.
Because we used the same PSF templates and photometric apertures as SEDS, 
we also used the same SEDS photometric corrections to correct the original,
PSF-fitted magnitudes to total magnitudes.  All S-CANDELS catalogs include these 
aperture corrections, which are given by Ashby et al.\ (2013; their Table~2).  
To avoid spurious sources, only objects detected in both bands 
were included in the catalogs.

\begin{figure}
\epsscale{1.2}
\includegraphics[bb=18 144 592 718,width=\columnwidth]{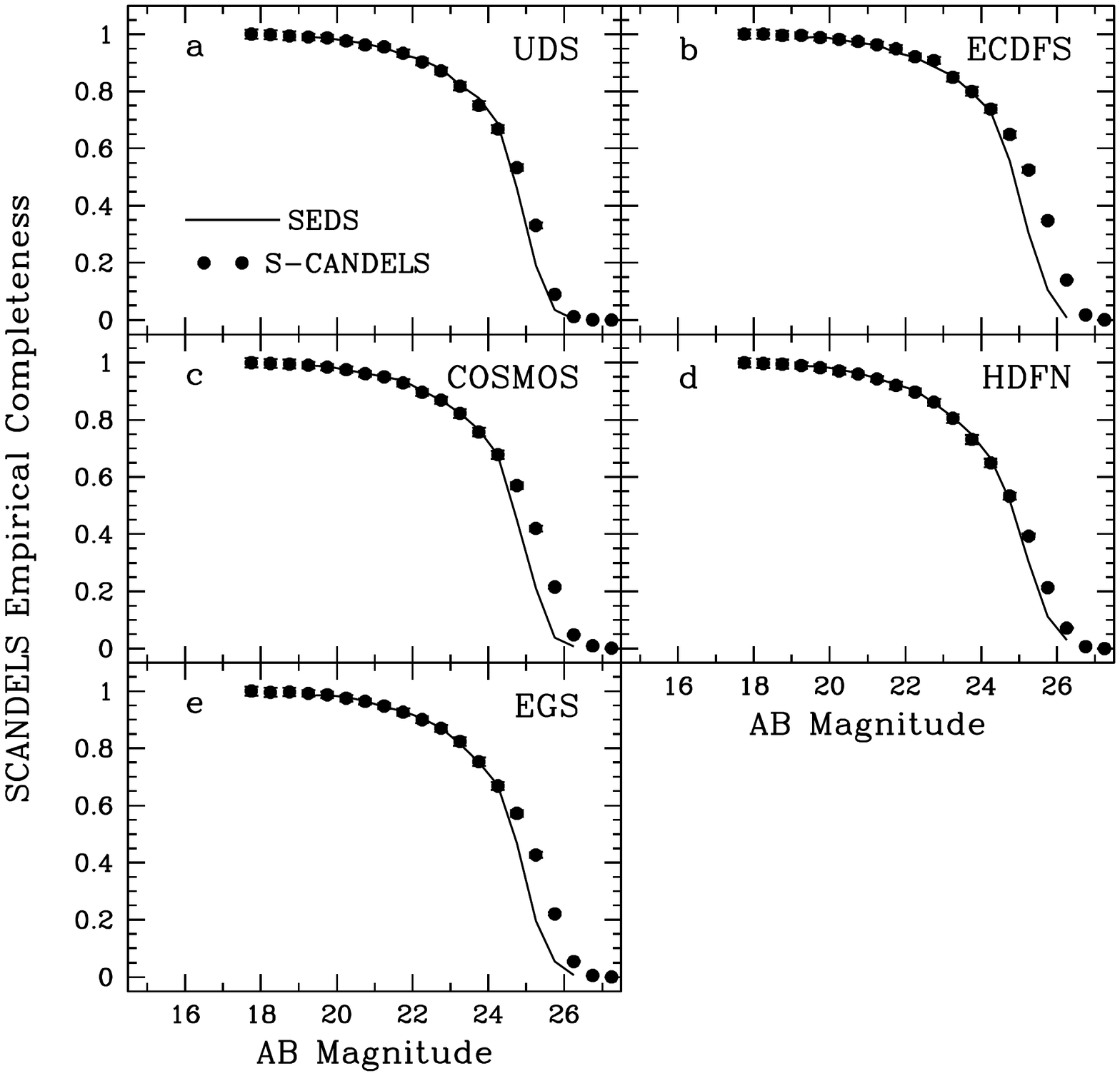}
\caption{ 
Completeness in the S-CANDELS fields estimated by Monte Carlo simulations 
as described in Section~\ref{ssec:extraction}.  Symbols indicate the completeness
measured in bins of width 0.5\,mag within the S-CANDELS fields shown in 
Figures~\ref{fig:UDS1} to \ref{fig:EGS2} where the depth of coverage was at least 
12\,hr; the mean
coverage was by design signficantly higher (Fig~\ref{fig:depths}).  
The completeness measured for SEDS, specifically coverage of at least 10\,ks 
total exposure time in the wider-but-shallower SEDS mosaics from Ashby et al. (2013), 
is indicated with the black line.  In both cases the values given are by-field averages.  
The completeness in any particular small region depends on the actual integration 
time achieved there and on the highly variable local source confusion.
\label{fig:comp}
}
\end{figure}

The completeness and reliability of S-CANDELS were assessed on a field-by-field
basis with the standard Monte Carlo approach, identical to that used for SEDS
(Ashby et al.\ 2013).
SEDS established, by matching CANDELS F160W sources to IRAC-detected
objects in COSMOS, that all IRAC sources fainter than 23\,AB mag are point 
sources at IRAC resolution.  This is true even for a majority of IRAC 
sources brighter than 23\,AB mag.  We therefore used only point sources in 
our completeness and depth simulations.  
We simultaneously inserted simulated sources 
in both the 3.6 and 4.5\,$\mu$m mosaics at identical locations.  For simplicity,
the simulated sources were created with [3.6]$-$[4.5]=0, i.e., 
no attempt was made to insert sources with a range of colors.  After the S-CANDELS
mosaics had been modified by inserting simulated sources, source identification
and photometry was performed in exactly the same way as for the unmodified
mosaics.  The completeness and magnitude bias were assessed by comparing the results
to the {\sl a priori} known input sources over the range of magnitudes seen
for the real sources.
The results are given in Table~\ref{tab:comp}
and shown in Figure~\ref{fig:comp}.  

S-CANDELS completeness is identical to that of SEDS for sources brighter than
about 24.5\,AB mag.  
For sources fainter than 24.5\,mag, however, S-CANDELS is significantly more 
complete than SEDS, recovering a larger but flux-dependent fraction of 
the simulated sources.  The improvement relative to SEDS ranges up to a 
factor of several, depending on the specific field and source 
magnitude.  Taken at face value, S-CANDELS reaches 50\% completeness 
at roughly [3.6]=[4.5]=25\,mag in all fields except the ECDFS, where
(because of the additional coverage from the ERS and IUDF programs), 
the 50\% completeness threshold is reached at 25.3\,mag.  
Users of S-CANDELS data are cautioned
that these are generalizations; the depths are variable 
across the S-CANDELS fields, and the completeness at any one location is 
a strong function of both the local source density and the total exposure time.

\begin{figure}
\epsscale{1.2}
\includegraphics[bb=18 144 592 718,width=\columnwidth]{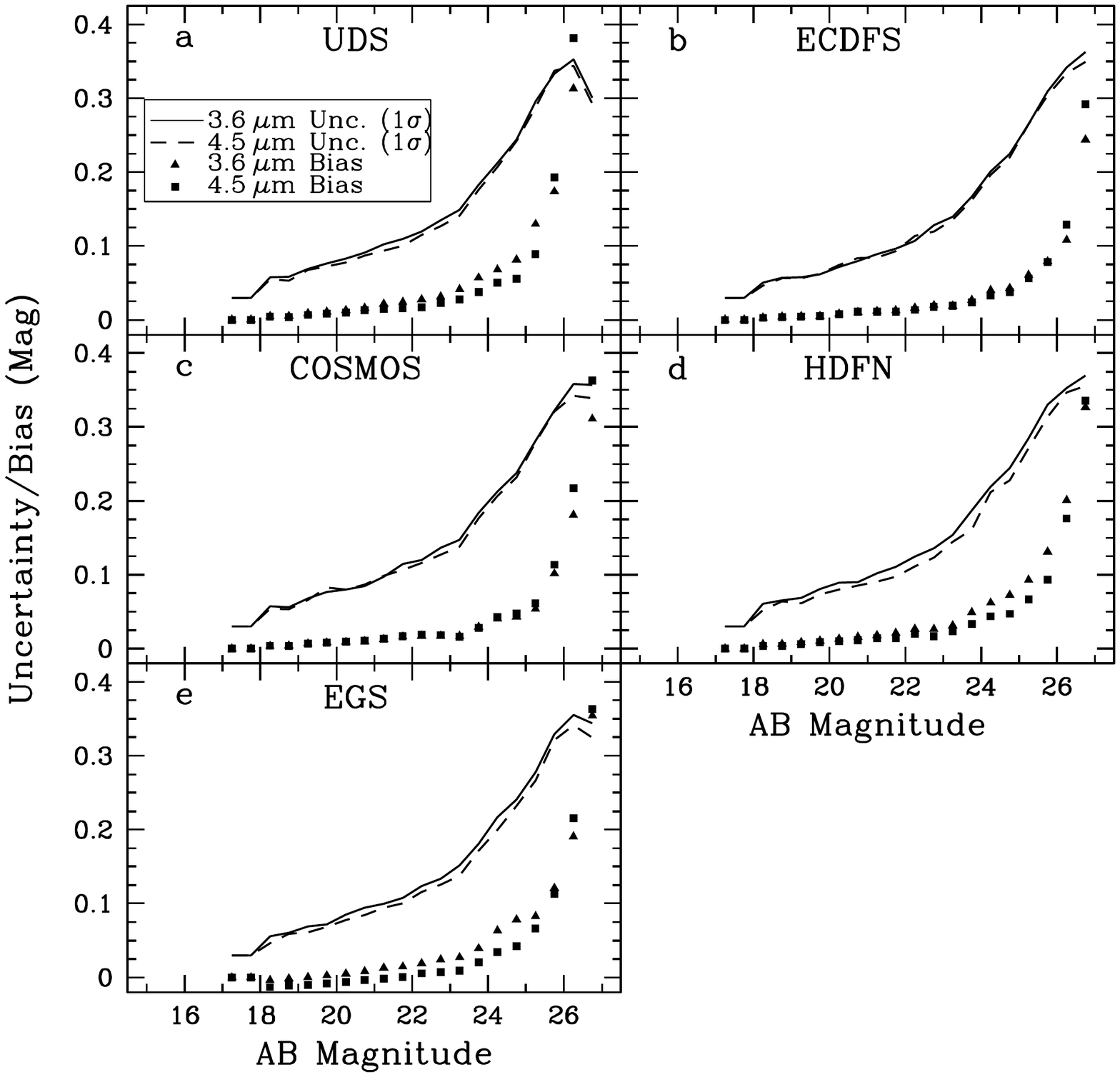}
\caption{ 
S-CANDELS measurement errors based on photometry of simulated sources
as described in Section~\ref{ssec:extraction}.  The symbols indicate the 
measurement bias 
as a function of apparent magnitude.  
Positive values indicate that 
measured values are brighter than the true values.  The S-CANDELS catalogs
have been corrected for this bias.
The solid and dashed lines indicate 
the 1$\sigma$ measurement uncertainty at 3.6 and 4.5\,$\mu$m, respectively.
The lower limit of 0.03\,mag on the measurement uncertainties reflects 
the uncertainty in the IRAC absolute calibration.
\label{fig:errors}
}
\end{figure}

As with SEDS, the S-CANDELS estimates of photometric error and bias 
were also based solely on the artificial source simulations, 
in order to account
for the impact of source confusion in these very deep mosaics.
The photometric uncertainties and biases are given respectively in
Tables~\ref{tab:errors} and \ref{tab:bias} 
for each of the S-CANDELS fields, and are shown in Figure~\ref{fig:errors}. 
The S-CANDELS photometric uncertainties are very close to those measured in
the shallower SEDS mosaics (Fig.~27 of Ashby et al. 2013).  This is discussed
in Section~\ref{ssec:confusion}.

As with SEDS, the measurement bias is relatively small for sources brighter 
than the 50\% completeness limit but grows rapidly at progressively fainter 
magnitudes (Table~\ref{tab:bias}).  This appears to confirm an interpretation 
in which faint sources are increasingly difficult to deblend from their 
neighbors.  The contamination of the photometric apertures by imperfectly 
subtracted brighter neighbors affects the photometry even though the 
measurements were made in source-subtracted residual images.  
The S-CANDELS catalogs have been corrected to remove 
the resulting average magnitude bias.

\subsection{Photometric Validation}
\label{sec:calibration}

We verified our astrometry by comparing the positions of extracted sources
to their counterparts in the references used by the CANDELS team.  We also
compared to the bright sources in the Two Micron All Sky Survey (2MASS; 
Skrutskie et al.\ 2006).  The results are shown in Table~\ref{tab:astrometry}.  
The S-CANDELS astrometry is consistent with previous work in the five
CANDELS fields.  The scatter measured for the positions of sources 
on the IRAC and non-IRAC catalogs is of order 0\farcs2, which is also
consistent with analogous measurements carried out in other fields, 
e.g., SEDS, the SSDF (Ashby et al.\ 2013b), and SDWFS (Ashby et al.\ 2009).

We verified our photometry by comparing to the measurements obtained
earlier in the shallower SEDS campaign, which were themselves already
validated against S-COSMOS (Sanders \etal 2007), SpUDS (version DR2), 
the EGS (Barmby \etal 2008), and SIMPLE (Damen \etal 2011).

\begin{figure}
\epsscale{1.25}
\includegraphics[bb=18 144 592 718,width=\columnwidth]{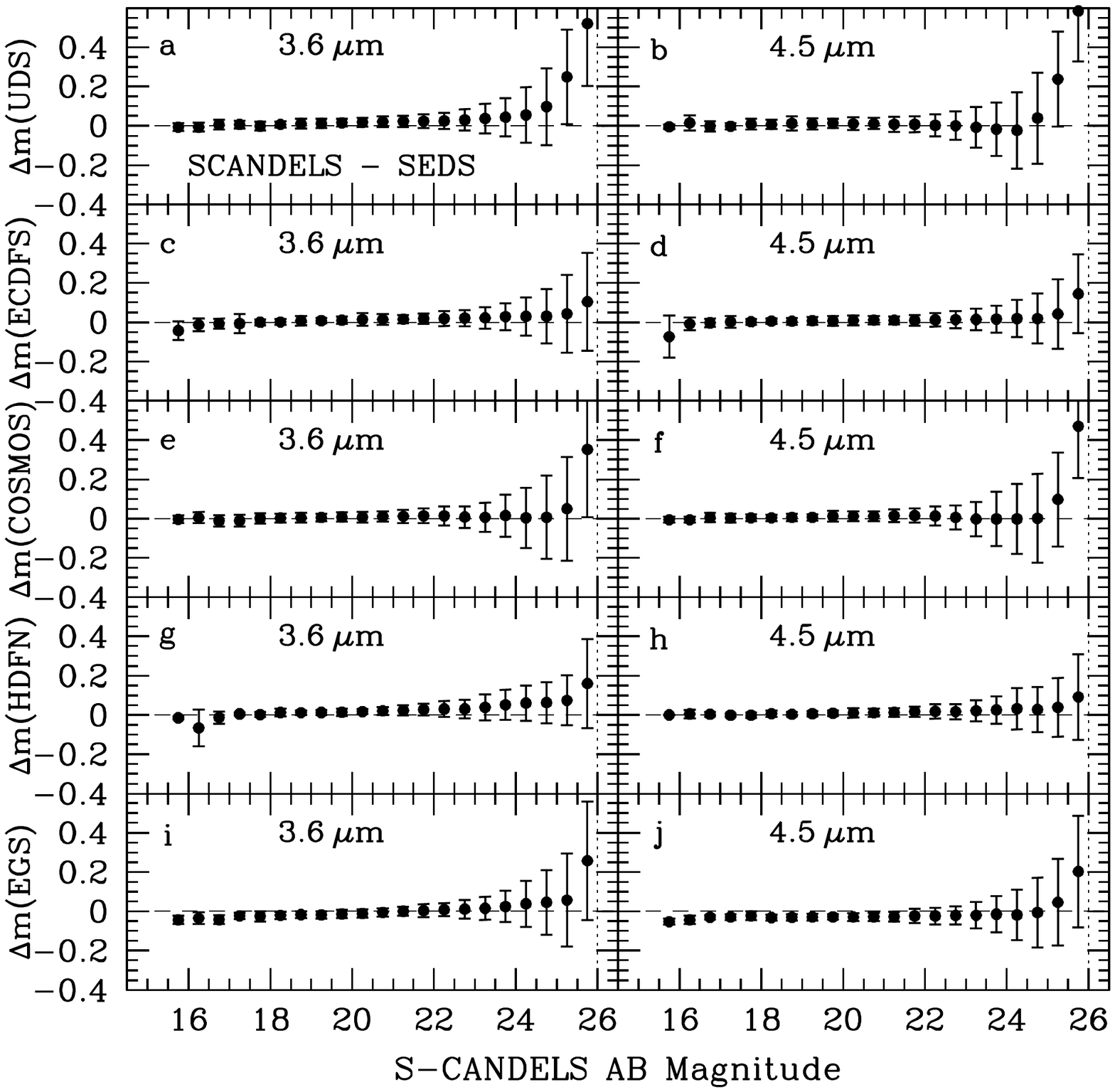}
\caption{
A comparison of S-CANDELS and SEDS photometry at 3.6 and 4.5\,$\mu$m.  
Symbols indicate the mean differences for measurements made in 2\farcs4 diameter
apertures within bins 0.5\,mag wide.  Positive-valued differences mean sources
appear brighter in SEDS than in S-CANDELS on average.
All error bars are 1$\sigma$.  Vertical 
dotted lines indicate the SEDS 3$\sigma$ sensitivity limits.  
The comparison was made after both photometric datasets were corrected to 
total magnitudes by compensating for empirically determined
aperture losses and biases.  
\label{fig:compare}
}
\end{figure}

To verify the S-CANDELS flux calibration, we matched the S-CANDELS catalogs 
to those from SEDS.  In all cases, the matching was
done using a $0\farcs5$ search radius, i.e., roughly twice the 
S-CANDELS $1\sigma$ astrometric uncertainty and one-third the
IRAC PSFs' FWHMs.  Only
SEDS sources brighter than 26\,AB mag (the SEDS 3$\sigma$ detection limit)
were used for the comparison.  Sources close to saturation 
($<$15.4\,mag in 200\,s exposures) were excluded.
The results are shown in Figure~\ref{fig:compare}.

\begin{deluxetable*}{cccccc}
\tabletypesize{\scriptsize}
\tablecolumns{6}
\tablewidth{0pc}
\tablecaption{Completeness in the S-CANDELS IRAC Catalogs\label{tab:comp}}
\tablehead{
AB Mag & UDS & ECDFS & COSMOS & HDFN & EGS \\
}
\startdata
 18.25 &   0.998$\pm$0.016 & 1.000$\pm$0.019 & 0.997$\pm$0.015 & 0.997$\pm$0.015 & 0.996$\pm$0.020 \\
 18.75 &   0.994$\pm$0.015 & 0.996$\pm$0.017 & 0.995$\pm$0.013 & 0.995$\pm$0.014 & 0.997$\pm$0.019 \\
 19.25 &   0.990$\pm$0.011 & 0.996$\pm$0.017 & 0.991$\pm$0.011 & 0.990$\pm$0.011 & 0.992$\pm$0.011 \\
 19.75 &   0.987$\pm$0.011 & 0.988$\pm$0.016 & 0.984$\pm$0.010 & 0.982$\pm$0.011 & 0.987$\pm$0.011 \\
 20.25 &   0.976$\pm$0.011 & 0.982$\pm$0.014 & 0.975$\pm$0.010 & 0.971$\pm$0.011 & 0.975$\pm$0.011 \\
 20.75 &   0.962$\pm$0.010 & 0.974$\pm$0.013 & 0.961$\pm$0.010 & 0.960$\pm$0.010 & 0.964$\pm$0.011 \\
 21.25 &   0.955$\pm$0.010 & 0.963$\pm$0.011 & 0.950$\pm$0.010 & 0.943$\pm$0.010 & 0.948$\pm$0.011 \\
 21.75 &   0.933$\pm$0.013 & 0.949$\pm$0.015 & 0.929$\pm$0.013 & 0.920$\pm$0.013 & 0.926$\pm$0.012 \\
 22.25 &   0.903$\pm$0.012 & 0.921$\pm$0.016 & 0.896$\pm$0.012 & 0.897$\pm$0.013 & 0.900$\pm$0.011 \\
 22.75 &   0.871$\pm$0.012 & 0.908$\pm$0.016 & 0.869$\pm$0.010 & 0.862$\pm$0.020 & 0.870$\pm$0.010 \\
 23.25 &   0.818$\pm$0.015 & 0.849$\pm$0.015 & 0.822$\pm$0.010 & 0.805$\pm$0.022 & 0.824$\pm$0.015 \\
 23.75 &   0.751$\pm$0.015 & 0.800$\pm$0.014 & 0.757$\pm$0.010 & 0.731$\pm$0.021 & 0.753$\pm$0.014 \\
 24.25 &   0.668$\pm$0.014 & 0.738$\pm$0.009 & 0.678$\pm$0.009 & 0.649$\pm$0.011 & 0.668$\pm$0.013 \\
 24.75 &   0.533$\pm$0.012 & 0.649$\pm$0.009 & 0.570$\pm$0.008 & 0.533$\pm$0.009 & 0.572$\pm$0.012 \\
 25.25 &   0.331$\pm$0.010 & 0.525$\pm$0.009 & 0.420$\pm$0.007 & 0.393$\pm$0.008 & 0.427$\pm$0.011 \\
 25.75 &   0.090$\pm$0.005 & 0.348$\pm$0.007 & 0.215$\pm$0.005 & 0.213$\pm$0.006 & 0.221$\pm$0.008 \\
 26.25 &   0.012$\pm$0.002 & 0.140$\pm$0.004 & 0.048$\pm$0.002 & 0.071$\pm$0.003 & 0.054$\pm$0.004 \\
 26.75 &   0.001$\pm$0.000 & 0.018$\pm$0.002 & 0.009$\pm$0.002 & 0.007$\pm$0.001 & 0.005$\pm$0.001 \\
 27.25 &   0.000$\pm$0.000 & 0.001$\pm$0.000 & 0.001$\pm$0.001 & 0.000$\pm$0.000 & 0.000$\pm$0.000 \\
\enddata
\tablecomments{Completeness estimates for the S-CANDELS fields.  The magnitudes correspond
to the centers of bins of width 0.5\,mag in which the completeness was estimated.  
The completeness is unity at brighter magnitudes than those listed.  These completeness estimates 
were made for sources detected in both IRAC bands.
}
\end{deluxetable*}

For sources brighter than 25\,mag in all fields, S-CANDELS photometry agrees
very well with SEDS.  There are a few exceptions.  In HDFN three of seven 
sources in the [3.6]=(16.0,16.5) bin differ by $\sim0.2$\,mag from SEDS, 
and two of five sources in the brightest 4.5\,$\mu$m ECDFS bin are discrepant
at a similar level.  All of the discrepant sources are bright point sources
(Milky Way stars).  The S-CANDELS photometry in the complementary IRAC band 
for these sources agrees with that from SEDS (Fig.~\ref{fig:compare}).  
Variability is therefore unlikely to be the issue.  
All discrepant sources lie in parts of 
the mosaics that combine SEDS and S-CANDELS exposures, and moreover 
the S-CANDELS residual images for these sources are markedly different than
those from SEDS.  This suggests that although the PSF fitting technique 
worked 
for the vast majority of S-CANDELS sources, it failed for these 
few objects, for reasons particular to the details of their immediate 
surroundings and the mechanics of {\tt StarFinder}.

The photometry for faint sources follows a more consistent pattern.  
Over a wide magnitude range in both SCANDELS bands the agreement 
between SCANDELS and SEDS is excellent.
In all five fields, however, as the SEDS 26\,mag sensitivity 
limit is approached, a bias becomes apparent in the sense that 
SEDS sources are systematically brighter than their S-CANDELS counterparts.
The bias is lowest overall in the HDFN ($\sim$0.1\,mag), 
and highest in the UDS ($\sim$0.5\,mag).

To better understand the reason for the faint-source bias, we inspected 
both the SEDS and S-CANDELS data (mosaics and catalogs) at the locations of 
the most problematic sources, i.e., those with discrepancies greater
than 0.5\,mag.  Apart from a tendency -- by no means universal -- to lie in the
outskirts of bright sources, the discrepant sources present no obvious 
common trait in the residual images.  They do not lie in regions with 
obvious background artifacts.  Indeed, the S-CANDELS and SEDS photometry 
of neighbors within $7\arcsec$ of discrepant sources agree within the 
uncertainties, with very few exceptions.  The discrepancies 
are therefore not attributable to issues with the background modeling.
The vast majority of discrepant sources also have the same
number of neighbors within $7\arcsec$ in both SEDS and S-CANDELS.  The 
problem therefore does not generally arise from the {\tt StarFinder} 
deblending procedure; the same numbers of sources lie in the peripheries
of the discrepant sources in both SEDS and S-CANDELS.  Finally, we compared
the coordinate offsets of matched SEDS and S-CANDELS sources.  We found no evidence
to suggest that the most discrepant sources were significantly spatially offset
in SEDS and S-CANDELS, relative to sources with consistent photometry.  
Inappropriate placement of the {\tt StarFinder} PSF centroids and apertures is 
therefore not likely to be the problem.

\begin{deluxetable}{cccccc}
\tabletypesize{\small}
\tabletypesize{\scriptsize}
\tablecolumns{5}
\tablewidth{0pc}
\tablecaption{Empirical Photometric Uncertainties for S-CANDELS\label{tab:errors}}
\tablehead{
\colhead{AB Mag} & \colhead{UDS}  & \colhead{ECDFS}      & \colhead{COSMOS}      & \colhead{HDFN} & \colhead{EGS} 
}
\startdata
\cutinhead{3.6\,$\mu$m}
 16.25 &  0.03     &  0.03    &  0.03 &  0.03 &  0.03 \\ 
 16.75 &  0.03     &  0.03    &  0.03 &  0.03 &  0.03 \\ 
 17.25 &  0.03     &  0.03    &  0.03 &  0.03 &  0.03 \\ 
 17.75 &  0.03     &  0.03    &  0.03 &  0.03 &  0.03 \\ 
 18.25 &  0.06     &  0.05    &  0.06 &  0.06 &  0.06 \\ 
 18.75 &  0.06     &  0.06    &  0.06 &  0.07 &  0.06 \\ 
 19.25 &  0.07     &  0.06    &  0.07 &  0.07 &  0.07 \\ 
 19.75 &  0.08     &  0.06    &  0.08 &  0.08 &  0.07 \\ 
 20.25 &  0.08     &  0.07    &  0.08 &  0.09 &  0.08 \\ 
 20.75 &  0.09     &  0.08    &  0.08 &  0.09 &  0.09 \\ 
 21.25 &  0.10     &  0.09    &  0.10 &  0.10 &  0.10 \\ 
 21.75 &  0.11     &  0.10    &  0.11 &  0.11 &  0.11 \\ 
 22.25 &  0.12     &  0.11    &  0.12 &  0.12 &  0.12 \\ 
 22.75 &  0.14     &  0.13    &  0.14 &  0.14 &  0.13 \\ 
 23.25 &  0.15     &  0.14    &  0.15 &  0.15 &  0.15 \\ 
 23.75 &  0.18     &  0.17    &  0.18 &  0.19 &  0.18 \\ 
 24.25 &  0.21     &  0.20    &  0.21 &  0.22 &  0.22 \\ 
 24.75 &  0.24     &  0.22    &  0.24 &  0.24 &  0.24 \\ 
 25.25 &  0.30     &  0.26    &  0.28 &  0.28 &  0.28 \\ 
 25.75 &  0.33     &  0.31    &  0.32 &  0.33 &  0.33 \\ 
 26.25 &  0.35     &  0.34    &  0.36 &  0.35 &  0.36 \\ 
\cutinhead{4.5\,$\mu$m}
 16.25 &  0.03     &  0.03    &  0.03 &  0.03 &  0.03 \\ 
 16.75 &  0.03     &  0.03    &  0.03 &  0.03 &  0.03 \\ 
 17.25 &  0.03     &  0.03    &  0.03 &  0.03 &  0.03 \\ 
 17.75 &  0.03     &  0.03    &  0.03 &  0.03 &  0.03 \\ 
 18.25 &  0.06     &  0.05    &  0.05 &  0.05 &  0.05 \\ 
 18.75 &  0.05     &  0.06    &  0.05 &  0.06 &  0.06 \\ 
 19.25 &  0.07     &  0.06    &  0.07 &  0.06 &  0.06 \\ 
 19.75 &  0.07     &  0.06    &  0.08 &  0.07 &  0.07 \\ 
 20.25 &  0.08     &  0.07    &  0.08 &  0.08 &  0.08 \\ 
 20.75 &  0.09     &  0.08    &  0.09 &  0.09 &  0.08 \\ 
 21.25 &  0.09     &  0.08    &  0.10 &  0.09 &  0.09 \\ 
 21.75 &  0.10     &  0.09    &  0.11 &  0.10 &  0.10 \\ 
 22.25 &  0.11     &  0.11    &  0.12 &  0.11 &  0.12 \\ 
 22.75 &  0.13     &  0.12    &  0.13 &  0.12 &  0.13 \\ 
 23.25 &  0.14     &  0.14    &  0.14 &  0.14 &  0.14 \\ 
 23.75 &  0.18     &  0.16    &  0.18 &  0.16 &  0.17 \\ 
 24.25 &  0.21     &  0.20    &  0.21 &  0.21 &  0.20 \\ 
 24.75 &  0.24     &  0.22    &  0.23 &  0.23 &  0.23 \\ 
 25.25 &  0.29     &  0.26    &  0.28 &  0.27 &  0.27 \\ 
 25.75 &  0.34     &  0.30    &  0.32 &  0.31 &  0.32 \\ 
 26.25 &  0.34     &  0.33    &  0.34 &  0.35 &  0.34 \\ 
\enddata
\tablecomments{Empirically determined S-CANDELS 1$\sigma$ photometric uncertainties 
(magnitudes) determined using the Monte Carlo simulations described in 
Section~\ref{ssec:extraction}.
An estimated 3\% systematic error in the IRAC flux calibration is
included and limits the uncertainties for bright sources.  Sources
brighter than 14.7 AB mag are saturated in S-CANDELS.
}
\end{deluxetable}

\begin{deluxetable}{cccccc}
\tabletypesize{\small}
\tabletypesize{\scriptsize}
\tablecolumns{5}
\tablewidth{0pc}
\tablecaption{Photometric Bias in S-CANDELS Catalogs\label{tab:bias}}
\tablehead{
\colhead{Mag} & \colhead{UDS}  & \colhead{ECDFS}      & \colhead{COSMOS}      & \colhead{HDFN} & \colhead{EGS} 
}
\startdata
\cutinhead{3.6\,$\mu$m}
 17.75 &  0.00     &  0.00    &  0.00 &  0.00 &  0.00 \\ 
 18.25 &  0.01     &  0.00    &  0.00 &  0.01 &  0.00 \\ 
 18.75 &  0.01     &  0.00    &  0.00 &  0.01 &  0.00 \\ 
 19.25 &  0.01     &  0.01    &  0.01 &  0.01 &  0.00 \\ 
 19.75 &  0.01     &  0.01    &  0.01 &  0.01 &  0.00 \\ 
 20.25 &  0.01     &  0.01    &  0.01 &  0.01 &  0.01 \\ 
 20.75 &  0.02     &  0.01    &  0.01 &  0.02 &  0.01 \\ 
 21.25 &  0.02     &  0.01    &  0.01 &  0.02 &  0.01 \\ 
 21.75 &  0.02     &  0.01    &  0.02 &  0.02 &  0.01 \\ 
 22.25 &  0.03     &  0.02    &  0.02 &  0.03 &  0.02 \\ 
 22.75 &  0.03     &  0.02    &  0.02 &  0.03 &  0.02 \\ 
 23.25 &  0.04     &  0.02    &  0.02 &  0.03 &  0.03 \\ 
 23.75 &  0.06     &  0.03    &  0.03 &  0.05 &  0.04 \\ 
 24.25 &  0.07     &  0.04    &  0.04 &  0.06 &  0.06 \\ 
 24.75 &  0.08     &  0.04    &  0.04 &  0.07 &  0.08 \\ 
 25.25 &  0.13     &  0.06    &  0.05 &  0.09 &  0.08 \\ 
 25.75 &  0.17     &  0.08    &  0.10 &  0.13 &  0.12 \\ 
 26.25 &  0.31     &  0.11    &  0.18 &  0.20 &  0.19 \\ 
\cutinhead{4.5\,$\mu$m}
 18.75 &  0.00     &  0.00    &  0.00 &  0.00 &  0.00 \\ 
 19.25 &  0.01     &  0.00    &  0.01 &  0.01 &  0.00 \\ 
 19.75 &  0.01     &  0.01    &  0.01 &  0.01 &  0.00 \\ 
 20.25 &  0.01     &  0.01    &  0.01 &  0.01 &  0.00 \\ 
 20.75 &  0.01     &  0.01    &  0.01 &  0.01 &  0.00 \\ 
 21.25 &  0.01     &  0.01    &  0.01 &  0.01 &  0.00 \\ 
 21.75 &  0.02     &  0.01    &  0.02 &  0.01 &  0.00 \\ 
 22.25 &  0.02     &  0.01    &  0.02 &  0.02 &  0.01 \\ 
 22.75 &  0.02     &  0.02    &  0.02 &  0.02 &  0.01 \\ 
 23.25 &  0.03     &  0.02    &  0.02 &  0.02 &  0.01 \\ 
 23.75 &  0.04     &  0.02    &  0.03 &  0.03 &  0.02 \\ 
 24.25 &  0.05     &  0.03    &  0.04 &  0.04 &  0.03 \\ 
 24.75 &  0.06     &  0.04    &  0.05 &  0.05 &  0.04 \\ 
 25.25 &  0.09     &  0.06    &  0.06 &  0.07 &  0.07 \\ 
 25.75 &  0.19     &  0.08    &  0.11 &  0.09 &  0.11 \\ 
 26.25 &  0.38     &  0.13    &  0.22 &  0.18 &  0.21 \\ 
\enddata
\tablecomments{Mean photometric bias in the S-CANDELS fields (magnitudes), determined 
empirically using the Monte Carlo simulations described in 
Section~\ref{ssec:extraction}.  The bias is zero for sources brighter than the
brightest magnitude listed in the Table.  The sense of the bias
is that artificial sources are measured to be brighter, on average, than they
were {\sl a priori} known to be, by the amounts listed.  These biases have 
already been corrected in the catalogs presented here.}
\end{deluxetable}

Having ruled out issues with offset coordinates, poor background estimation, 
and deblending of different numbers of neighbors, we tentatively attribute the 
faint-source bias to flux boosting by very-low-level cosmic rays that are not 
efficiently rejected at SEDS depths.  With the factor-of-four greater number 
of exposures available to S-CANDELS, there is statistical power to 
reject faint outliers that cannot be ruled out at SEDS depths.  
This hypothesis is consistent with the fact that the bias is seen to be 
most pronounced in the faintest two SEDS magnitude bins, and is of roughly 
the same size as the SEDS uncertainties themselves.
The S-CANDELS magnitudes show similar bias but only for sources
roughly 0.5\,mag fainter than for SEDS, so we cannot rule out an
analogous effect in the faintest S-CANDELS bins (cf. Table~\ref{tab:bias}, Ashby 
et al.\ 2013, Table 5).

\begin{deluxetable*}{crrcc}
\tabletypesize{\scriptsize}
\tablecaption{S-CANDELS Astrometric Offsets\label{tab:astrometry}}
\tablewidth{0pt}
\tablehead{
\colhead{Field}  & \colhead{$\Delta$RA} & \colhead{$\Delta$Dec} & \colhead{Total} & \colhead{Coordinate}\\
                 & \colhead{(arcsec)} & \colhead{(arcsec)} & \colhead{(arcsec)} &  \colhead{Reference}
}
\startdata
        & \multicolumn{3}{c}{Relative to CANDELS}\\
UDS 	&  $0.00\pm0.13$ & $-0.02\pm0.14$ & $0.14\pm0.10$ & UKIDSS DR8 (Lawrence et al.\ 2007)    \\
ECDFS 	&  $0.02\pm0.16$ & $-0.19\pm0.15$ & $0.27\pm0.10$ & GOODS r2.0z (Giavalisco et al.\ 2004) \\
COSMOS 	&  $0.02\pm0.15$ & $ 0.04\pm0.16$ & $0.19\pm0.11$ & COSMOS v2.0 (Koekemoer et al.\ 2007)  \\
HDFN 	&  $0.12\pm0.27$ & $ 0.01\pm0.19$ & $0.30\pm0.17$ & GOODS r2.0z (Giavalisco et al.\ 2004) \\
EGS 	&  $0.03\pm0.14$ & $-0.02\pm0.16$ & $0.19\pm0.20$ & Lotz et al. (2008) \\
        & \multicolumn{3}{c}{Relative to 2MASS}\\
UDS 	&  $ 0.00\pm0.18$  & $-0.03\pm0.19$ & $0.23\pm0.12$ & Skrutskie et al.\ (2006) \\
ECDFS 	&  $-0.01\pm0.16$  & $0.03\pm0.18$  & $0.22\pm0.11$ & " \\
COSMOS 	&  $-0.02\pm0.17$  & $0.00\pm0.17$  & $0.22\pm0.11$ & " \\
HDFN 	&  $-0.03\pm0.16$  & $0.00\pm0.16$  & $0.20\pm0.11$ & " \\
EGS 	&   $0.03\pm0.18$  & $0.00\pm0.16$  & $0.21\pm0.12$ & " \\
\enddata
\tablecomments{Mean coordinate offsets measured for S-CANDELS
relative to astrometric references.  The upper half of the Table
compares the S-CANDELS IRAC source positions to
the astrometric references adopted by CANDELS.  
The bottom half of the Table compares the S-CANDELS
source positions to 2MASS.
Total offsets refer to the mean absolute offsets.
The stated uncertainties are the standard
deviations of the offset distributions for matched sources.
}
\end{deluxetable*}

\section{Discussion}
\label{sec:discussion}
\subsection{Number Counts}
\label{ssec:counts}

\begin{figure*}
\epsscale{0.73}
\includegraphics[bb=18 144 592 718,width=\textwidth]{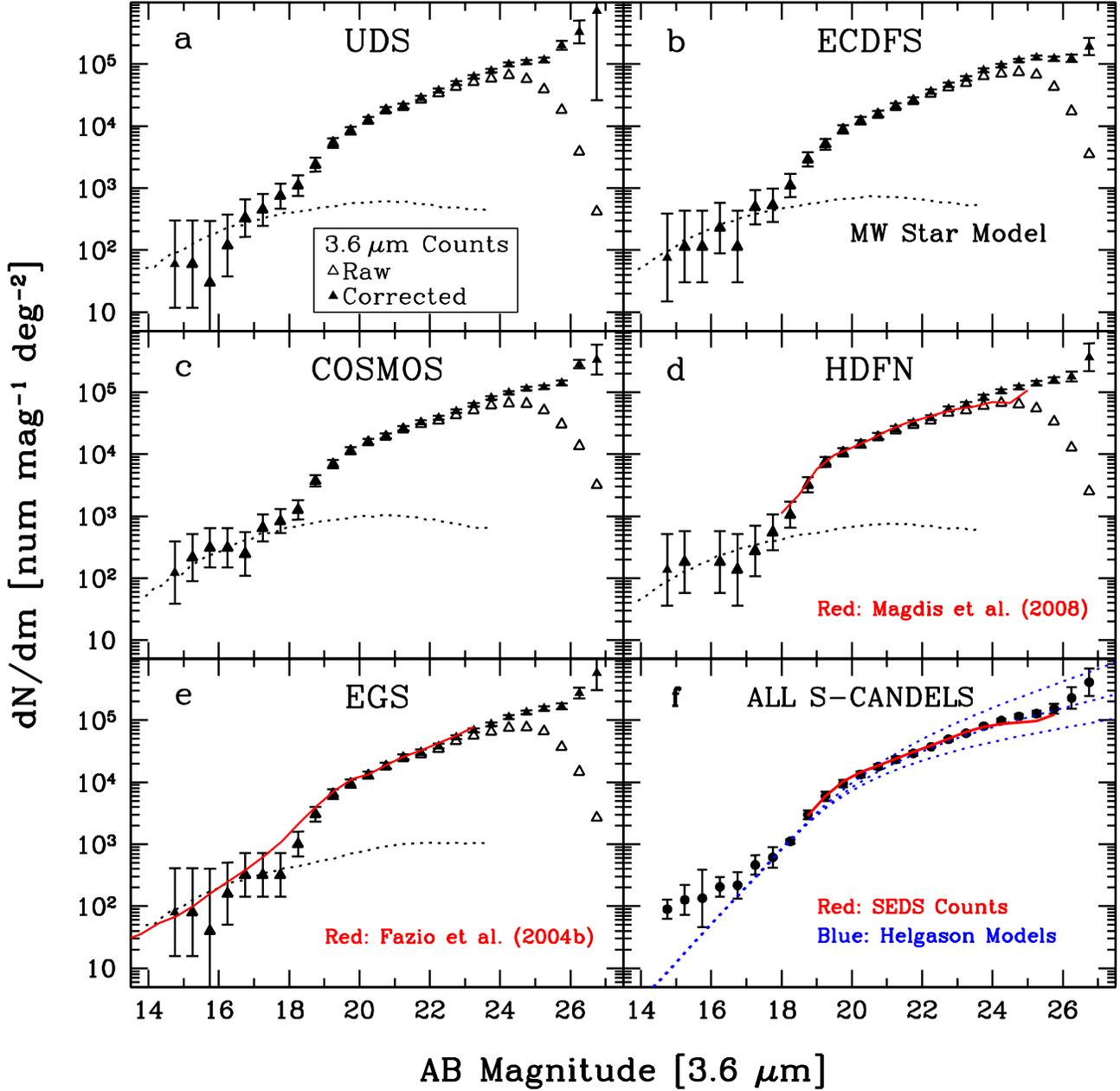}
\caption{
Differential 3.6\,$\mu$m counts in the five S-CANDELS fields.  Open symbols show
the raw counts, while solid symbols indicate the counts corrected for incompleteness
on the basis of simulated detections as described in Section~\ref{ssec:extraction}.  
Error bars represent the Poisson statistics only in panels a-e.  The solid red lines
in panels d and e indicate the incompleteness-corrected counts measured in the HDFN
by Magdis \etal (2008) and in the EGS by Fazio \etal (2004b), respectively.  
The dotted lines in panels a--e show the expected counts arising from 
Milky Way stars, based on the DIRBE 
Faint Source Model at 3.5\,$\mu$m (Arendt \etal 1998; Wainscoat et al.\ 1992; 
Cohen \etal 1993, 1994, 1995).  Panel f shows the area-weighted mean counts 
for all of S-CANDELS together with predicted counts from Helgason et al.\ (2012).
The upper and lower blue dotted lines indicate the Helgason et al.\ high-faint-end
and low-faint-end luminosity function models, i.e., models in which the slopes of
the faint end of the luminosity functions were respectively set to $\alpha=-1.2$
and $-0.8$.  The middle blue dashed line indicates the 
so-called `default' model, obtained by  
averaging the high-faint-end and low-faint-end models.  The SEDS source counts 
are shown in red.
\label{fig:counts1}
}
\end{figure*}

\begin{figure*}
\epsscale{0.83}
\includegraphics[bb=18 144 592 718,width=\textwidth]{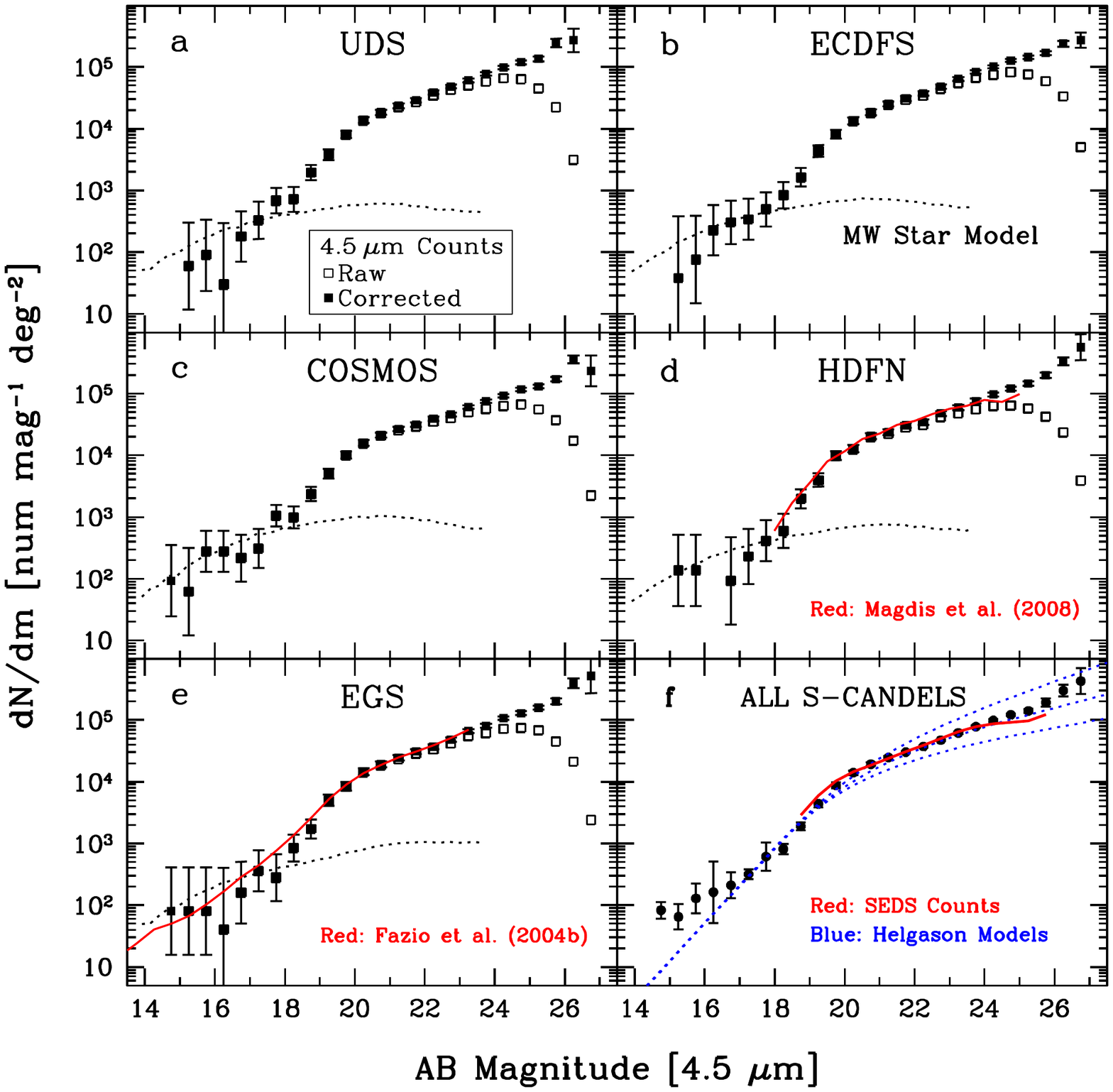}
\caption{
Differential counts in the five S-CANDELS fields at 4.5\,$\mu$m.  The meanings of the 
symbols are the same as in Fig.~\ref{fig:counts1}.  The Milky Way Star Models shown 
are those for DIRBE at 4.9\,$\mu$m, from Arendt \etal (1998), Wainscoat et al.\ (1992),
and Cohen \etal (1993, 1994, 1995).
\label{fig:counts2}
}
\end{figure*}

S-CANDELS detects roughly 135,000 sources in the combined 0.16\,deg$^2$ 
area covered by the five fields in the survey.  
Figures~\ref{fig:counts1} and \ref{fig:counts2} and
Table~\ref{tab:counts} present the resulting differential source counts
along with Milky Way star counts estimated from the Arendt \etal (1998) model for the 
S-CANDELS lines-of-sight.  

The S-CANDELS counts rely on completeness corrections that are based on simulated
sources with zero color, i.e., $[3.6]-[4.5]=0$.  At faint levels they could therefore 
in principle suffer from subtle systematic effects, because real sources 
span a range of colors (Figure~\ref{fig:mhist}).  Our simulations do not 
fully account for faint, blue 3.6\,$\mu$m sources, which would tend to elude 
detection in the 4.5\,$\mu$m band.  Faint, red 4.5\,$\mu$m sources 
would be under-counted for the same reason.  However, these systematic 
effects are unlikely to severely bias the S-CANDELS counts.  The real IRAC color
distribution peaks at $[3.6]-[4.5]=0$, and the vast majority ($\sim80$\%) 
have colors within 0.4\,mag of the peak (Figure~\ref{fig:mhist}),
even down to faint levels.  Moreover, the area-weighted mean S-CANDELS 
counts (Figs.~\ref{fig:counts1}f and \ref{fig:counts2}f)
show very close agreement with SEDS over the full range of comparison.  For
$[3.6]=[4.5]>20$\,mag, the counts show no significant deviations from those
found for SEDS, suggesting the SEDS completeness corrections were accurate.
S-CANDELS uses the same techniques, so its completeness
corrections should be similarly robust.

At levels brighter than roughly 18\,AB mag 
in both S-CANDELS bands and in every field, the IRAC counts are consistent
with the star count models.  SEDS contains relatively few galaxies brighter
than 18\,mag.  The vast majority of sources fainter than 18\,mag, however, 
are galaxies: the contributions of Milky Way stars 
to the faint counts are negligible.  

Helgason et al.\ (2012) modeled galaxy counts using an ensemble of
galaxy luminosity functions assembled from deep multiband observations.  
They then used this ensemble to predict faint galaxy counts in
several passbands, including 3.6 and 4.5\,$\mu$m.  
They used existing counts to constrain the faint-end slopes
of their luminosity functions.  Specifically, only a limited
range of faint-end slopes, corresponding to a range of acceptable
values for the parameter $\alpha$, was found to be consistent
with existing counts.  That range extends from their so-called
high-faint-end, with $\alpha=-1.2$, to the low-faint-end ($\alpha=-0.8$).
They considered also a `default' model that averages these two cases.
The IRAC
counts closely follow the `default' model all the way down to
[3.6]=[4.5]=26\,mag (Figures~\ref{fig:counts1}f and \ref{fig:counts2}f).  
At fainter levels, the counts depart upward
in the direction of the high-faint-end scenario.  This may not be
real, because it occurs at magnitudes where the completeness correction is 
largest, magnifying any small systematic errors that might be present 
in the counts.  It is also consistent with the possibility that faint
sources undergo flux boosting, as described in the preceding Section.
What can be said
with confidence is that the Helgason et al.\ (2012) models work
very well down to very faint levels.  More sensitive observations 
that can overcome the source confusion seen in the
IRAC mosaics (e.g., imaging with the {\sl James Webb Space Telescope} 
or {\sl WISH (The Wide-Field Imaging Surveyor for High Redshift}) will be necessary 
to confirm this picture for the faintest IRAC-detected sources.

\subsection{Source Confusion}
\label{ssec:confusion}
 
For sources brighter than 24.5\,mag, the S-CANDELS source detection fraction 
is not significantly better than that of SEDS despite a factor-of-four
improvement in overall integration time.  Sources at 24.5\,mag or brighter
lie well above even the SEDS detection threshold.  Sensitivity
alone is therefore not limiting the bright-source detection by IRAC in
this regime.  
Moreover, the empirically determined S-CANDELS 
photometric uncertainties for bright sources are very similar to those for SEDS.  
We suggest that source confusion is the dominant contributor to the 
photometric uncertainties for magnitudes $<24.5$\,mag.  Deeper IRAC 
observations alone will not improve the detection fraction or the
photometric uncertainties for such bright sources.

For fainter sources, the picture is more nuanced.  Source confusion
is undoubtedly a factor, as evidenced by the similarity 
of SEDS and S-CANDELS uncertainties down to the limits of the surveys.
However, for the deeper S-CANDELS, the detection fraction at $>24.5$\,mag 
is up to factors of several larger.  Inspection
of the respective catalogs revealed that the majority
of the faint S-CANDELS sources {\sl not detected by SEDS} lie in 
relatively source-free portions of the fields.  It is in precisely these
places that the improvement in sensitivity can be effective at identifying
faint objects by decreasing the background shot noise.  

One way to better understand the impact of source confusion on deep 
IRAC imaging is to quantify the available source-free area that will
yield additional IRAC detections when imaged more deeply.
A conservative estimate of this area is that in which detected sources
contribute less surface brightness than the surface brightness noise
level $\sigma$.  We estimated $\sigma$ for both SEDS and S-CANDELS
using the residual images, i.e., after removing detected sources, and
allowing for the effect of correlated noise.\footnote{Each mosaic pixel
is one-fourth of an IRAC pixel, so the true noise is double the 
standard deviation measured in the mosaic.  Surface brightness due to
known sources was measured on the model mosaics, which by construction
include the contributions of all detected sources and nothing else.}
By this definition, the source-free areas in 12\,hr SEDS integrations
are $\sim40$\% and $\sim50$\% in the 3.6 and 4.5\,$\mu$m bands, 
respectively.  The fractions that remain free in the 50\,hr S-CANDELS 
mosaics are smaller, $\sim30$\% and $\sim40$\% at 3.6 and 4.5\,$\mu$m, 
respectively.  In other words, of order half the SEDS area and one-third
the S-CANDELS area is effectively clear.  Within those areas, 
integrating longer to reduce the shot noise can improve the detection
statistics, and we see the results in the increased S-CANDELS completeness
(relative to SEDS) for sources fainter than 24.5\,mag.

In summary, source confusion does play a role for faint sources, but
the much deeper S-CANDELS program nonetheless detects a significantly 
greater fraction of such sources.  Somewhat counter-intuitively, the 
faint IRAC sources are not as rigidly limited by source confusion 
as the bright ones.

\subsection{IRAC Color Distribution}

\begin{figure}
\includegraphics[bb=40 370 380 700, width=\columnwidth]{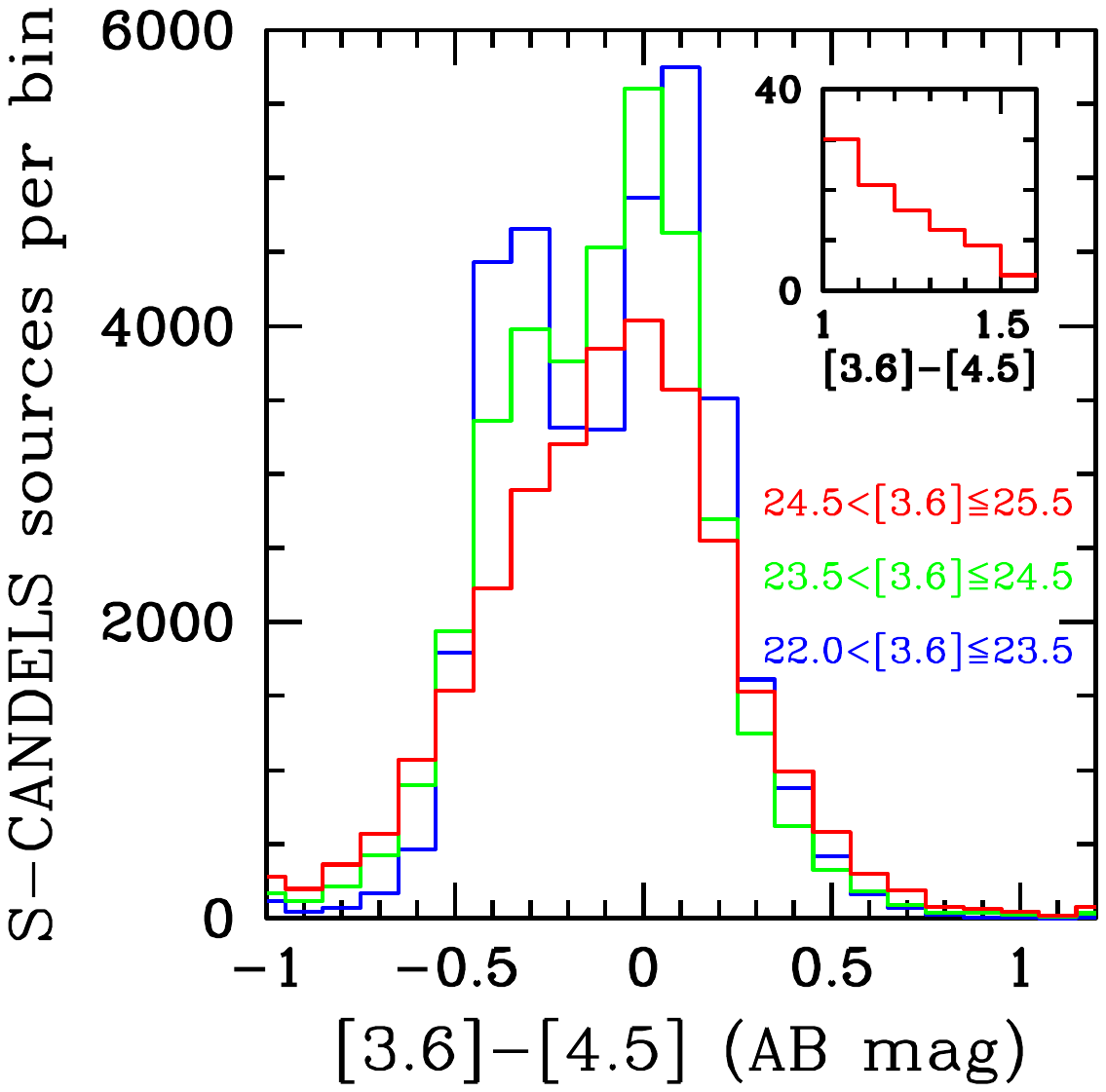}
\caption{Histograms of S-CANDELS sources in three magnitude bins as
indicated by colors.  The inset shows the distribution of the reddest
sources in the $24.5<[3.6]<25.5$ magnitude range.  There are 118
sources in this range with $[3.6]-[4.5]>1$ of which 27 have
$[3.6]-[4.5]>1.6$.  Given the photometric uncertainties (Fig.~16),
the apparent red colors may be due to photometric errors.
\label{fig:mhist}
}
\end{figure}

The IRAC colors of the sources give clues to their redshifts and 
luminosities.  For example, Sorba \& Sawicki (2010) and Barro et al.\ (2011) 
showed that the $[3.6]-[4.5]$ color is a useful photometric redshift 
indicator, especially for
separating galaxies at $z\la1.3$ from those at $z\ga1.5$.
Ashby et al.\ (2013) showed (their Fig.~31) that the observed color
distribution is in fact bimodal for $[3.6]<23.5$ and that the red
peak grows relative to the blue one as fainter sources are considered.
Figure~\ref{fig:mhist} shows the same trend for the fainter sources
observed in S-CANDELS.  Smaller color uncertainties than in SEDS give
a hint of bimodality for $23.5<[3.6]\le24.5$, but fainter
sources show a single peak.  Sources with blue colors are still
present, but their proportion is significantly smaller than at
brighter magnitudes.  The effect of increasing uncertainty in the
photometry of the faintest sources (Fig.~\ref{fig:errors}), and
likewise in their colors, is visible as a broadening
of the wings of their histogram.  The faintest sources plotted, at
$[3.6]\approx25$, correspond to $\sim$5~mag fainter than L$^*$ at $z=1.2$
or $\sim$3~mag fainter than L$^*$ at $z=2.9$ (based on
luminosity functions given by \citeauthor{Helgason2012}).  At $z>3$ 
the galaxy space density decreases approximately exponentially, and 
such galaxies will constitute only a small fraction of the sample.
Therefore most of the faint sources are likely to be galaxies a few
magnitudes fainter than L$^*$ at $z=1.2$--3.

\begin{deluxetable*}{c cc cc cc cc cc|cc}
\tabletypesize{\small}
\tabletypesize{\scriptsize}
\tablecolumns{5}
\tablewidth{0pc}
\tablecaption{S-CANDELS IRAC Number Counts\label{tab:counts}}
\tablehead{
\colhead{Mag}  & \multicolumn{2}{c}{UDS}  & \multicolumn{2}{c}{ECDFS}      & \multicolumn{2}{c}{COSMOS}      & \multicolumn{2}{c}{HDFN} &
\multicolumn{2}{c}{EGS} & \multicolumn{2}{c}{Total} \\
\colhead{(AB)} & \colhead{Counts} & \colhead{Unc.} 
& \colhead{Counts} & \colhead{Unc.} 
& \colhead{Counts} & \colhead{Unc.} 
& \colhead{Counts} & \colhead{Unc.} 
& \colhead{Counts} & \colhead{Unc.} 
& \colhead{Counts} & \colhead{Unc.}
}
\startdata
\cutinhead{3.6\,$\mu$m}
14.75 &  1.77 &   0.71 &  1.88 &   0.71 &  2.09 &   0.50 &  2.14 &   0.58 &  1.90 &   0.71 &  1.95 &   0.15 \\ 
15.25 &  1.77 &   0.71 &  2.06 &   0.58 &  2.33 &   0.38 &  2.26 &   0.50 &  1.90 &   0.71 &  2.10 &   0.24 \\ 
15.75 &  1.47 &   1.00 &  2.06 &   0.58 &  2.49 &   0.32 &  2.26 &   0.50 &  1.60 &   1.00 &  2.13 &   0.46 \\ 
16.25 &  2.08 &   0.50 &  2.36 &   0.41 &  2.49 &   0.32 &  2.26 &   0.50 &  2.20 &   0.50 &  2.31 &   0.16 \\ 
16.75 &  2.51 &   0.30 &  2.06 &   0.58 &  2.39 &   0.35 &  2.14 &   0.58 &  2.50 &   0.35 &  2.34 &   0.21 \\ 
17.25 &  2.65 &   0.26 &  2.69 &   0.28 &  2.81 &   0.22 &  2.44 &   0.41 &  2.50 &   0.35 &  2.67 &   0.16 \\ 
17.75 &  2.87 &   0.20 &  2.73 &   0.27 &  2.92 &   0.19 &  2.74 &   0.29 &  2.50 &   0.35 &  2.79 &   0.17 \\ 
18.25 &  3.04 &   0.17 &  3.04 &   0.19 &  3.10 &   0.16 &  3.02 &   0.21 &  3.00 &   0.20 &  3.05 &   0.04 \\ 
18.75 &  3.38 &   0.11 &  3.46 &   0.12 &  3.57 &   0.09 &  3.51 &   0.12 &  3.48 &   0.12 &  3.48 &   0.07 \\ 
19.25 &  3.73 &   0.08 &  3.70 &   0.09 &  3.84 &   0.07 &  3.87 &   0.08 &  3.81 &   0.08 &  3.78 &   0.07 \\ 
19.75 &  3.92 &   0.06 &  3.94 &   0.07 &  4.06 &   0.05 &  4.03 &   0.07 &  3.98 &   0.07 &  3.98 &   0.06 \\ 
20.25 &  4.10 &   0.05 &  4.09 &   0.06 &  4.20 &   0.05 &  4.17 &   0.06 &  4.12 &   0.06 &  4.13 &   0.05 \\ 
20.75 &  4.26 &   0.04 &  4.20 &   0.05 &  4.29 &   0.04 &  4.29 &   0.05 &  4.27 &   0.05 &  4.25 &   0.04 \\ 
21.25 &  4.32 &   0.04 &  4.33 &   0.04 &  4.42 &   0.04 &  4.41 &   0.04 &  4.41 &   0.04 &  4.37 &   0.05 \\ 
21.75 &  4.45 &   0.04 &  4.42 &   0.04 &  4.52 &   0.04 &  4.50 &   0.04 &  4.49 &   0.04 &  4.47 &   0.04 \\ 
22.25 &  4.57 &   0.03 &  4.55 &   0.04 &  4.59 &   0.03 &  4.59 &   0.04 &  4.58 &   0.04 &  4.57 &   0.02 \\ 
22.75 &  4.69 &   0.03 &  4.67 &   0.03 &  4.69 &   0.03 &  4.73 &   0.04 &  4.73 &   0.03 &  4.69 &   0.03 \\ 
23.25 &  4.79 &   0.03 &  4.77 &   0.03 &  4.79 &   0.03 &  4.80 &   0.04 &  4.83 &   0.03 &  4.79 &   0.03 \\ 
23.75 &  4.89 &   0.03 &  4.90 &   0.03 &  4.91 &   0.03 &  4.92 &   0.04 &  4.93 &   0.03 &  4.90 &   0.02 \\ 
24.25 &  4.99 &   0.03 &  4.98 &   0.03 &  4.99 &   0.03 &  5.01 &   0.03 &  5.05 &   0.03 &  5.00 &   0.03 \\ 
24.75 &  5.03 &   0.03 &  5.06 &   0.03 &  5.06 &   0.03 &  5.08 &   0.03 &  5.13 &   0.03 &  5.06 &   0.04 \\ 
25.25 &  5.07 &   0.04 &  5.11 &   0.03 &  5.08 &   0.03 &  5.14 &   0.04 &  5.19 &   0.04 &  5.11 &   0.05 \\ 
25.75 &  5.30 &   0.07 &  5.09 &   0.04 &  5.15 &   0.04 &  5.19 &   0.05 &  5.22 &   0.05 &  5.18 &   0.08 \\ 
26.25 &  5.52 &   0.18 &  5.09 &   0.06 &  5.45 &   0.07 &  5.25 &   0.08 &  5.44 &   0.09 &  5.36 &   0.18 \\ 
26.75 &  5.86 &   1.44 &  5.28 &   0.14 &  5.52 &   0.24 &  5.56 &   0.23 &  5.76 &   0.28 &  5.61 &   0.22 \\ 
\cutinhead{4.5\,$\mu$m}
14.75 &\nodata&\nodata &\nodata&\nodata &  1.97 &   0.58 &\nodata&\nodata &  1.90 &   0.71 &  1.92 &   0.14 \\ 
15.25 &  1.77 &   0.71 &  1.58 &   1.00 &  1.79 &   0.71 &  2.14 &   0.58 &  1.90 &   0.71 &  1.81 &   0.21 \\ 
15.75 &  1.95 &   0.58 &  1.88 &   0.71 &  2.44 &   0.33 &  2.14 &   0.58 &  1.90 &   0.71 &  2.11 &   0.24 \\ 
16.25 &  1.47 &   1.00 &  2.36 &   0.41 &  2.44 &   0.33 &\nodata&\nodata &  1.60 &   1.00 &  2.21 &   0.50 \\ 
16.75 &  2.25 &   0.41 &  2.48 &   0.35 &  2.33 &   0.38 &  1.96 &   0.71 &  2.20 &   0.50 &  2.32 &   0.21 \\ 
17.25 &  2.51 &   0.30 &  2.53 &   0.33 &  2.49 &   0.32 &  2.36 &   0.45 &  2.56 &   0.33 &  2.50 &   0.08 \\ 
17.75 &  2.83 &   0.21 &  2.69 &   0.28 &  3.02 &   0.17 &  2.62 &   0.33 &  2.45 &   0.38 &  2.79 &   0.23 \\ 
18.25 &  2.85 &   0.20 &  2.92 &   0.21 &  3.00 &   0.18 &  2.78 &   0.28 &  2.93 &   0.22 &  2.91 &   0.08 \\ 
18.75 &  3.29 &   0.12 &  3.21 &   0.15 &  3.37 &   0.12 &  3.30 &   0.15 &  3.24 &   0.15 &  3.28 &   0.06 \\ 
19.25 &  3.58 &   0.09 &  3.64 &   0.10 &  3.70 &   0.08 &  3.60 &   0.11 &  3.70 &   0.09 &  3.64 &   0.06 \\ 
19.75 &  3.90 &   0.06 &  3.91 &   0.07 &  4.00 &   0.06 &  4.00 &   0.07 &  3.93 &   0.07 &  3.94 &   0.05 \\ 
20.25 &  4.14 &   0.05 &  4.13 &   0.06 &  4.20 &   0.05 &  4.11 &   0.06 &  4.17 &   0.05 &  4.15 &   0.04 \\ 
20.75 &  4.27 &   0.04 &  4.27 &   0.05 &  4.33 &   0.04 &  4.31 &   0.05 &  4.28 &   0.05 &  4.29 &   0.03 \\ 
21.25 &  4.37 &   0.04 &  4.40 &   0.04 &  4.42 &   0.04 &  4.37 &   0.05 &  4.39 &   0.04 &  4.39 &   0.02 \\ 
21.75 &  4.46 &   0.04 &  4.49 &   0.04 &  4.50 &   0.04 &  4.48 &   0.04 &  4.49 &   0.04 &  4.48 &   0.02 \\ 
22.25 &  4.59 &   0.03 &  4.57 &   0.04 &  4.59 &   0.03 &  4.53 &   0.04 &  4.57 &   0.04 &  4.57 &   0.02 \\ 
22.75 &  4.68 &   0.03 &  4.68 &   0.03 &  4.66 &   0.03 &  4.68 &   0.04 &  4.68 &   0.03 &  4.67 &   0.01 \\ 
23.25 &  4.78 &   0.03 &  4.81 &   0.03 &  4.78 &   0.03 &  4.77 &   0.04 &  4.82 &   0.03 &  4.79 &   0.02 \\ 
23.75 &  4.89 &   0.03 &  4.92 &   0.03 &  4.87 &   0.03 &  4.88 &   0.04 &  4.91 &   0.03 &  4.89 &   0.02 \\ 
24.25 &  4.99 &   0.03 &  5.00 &   0.03 &  4.96 &   0.03 &  4.99 &   0.03 &  5.03 &   0.03 &  4.99 &   0.02 \\ 
24.75 &  5.08 &   0.03 &  5.10 &   0.03 &  5.07 &   0.03 &  5.08 &   0.03 &  5.11 &   0.03 &  5.09 &   0.02 \\ 
25.25 &  5.13 &   0.04 &  5.16 &   0.03 &  5.12 &   0.03 &  5.16 &   0.04 &  5.20 &   0.04 &  5.15 &   0.03 \\ 
25.75 &  5.39 &   0.07 &  5.23 &   0.03 &  5.23 &   0.04 &  5.29 &   0.04 &  5.30 &   0.04 &  5.28 &   0.07 \\ 
26.25 &  5.43 &   0.19 &  5.38 &   0.05 &  5.55 &   0.07 &  5.52 &   0.07 &  5.60 &   0.08 &  5.48 &   0.09 \\ 
26.75 &\nodata&\nodata &  5.44 &   0.12 &  5.37 &   0.25 &  5.75 &   0.21 &  5.71 &   0.28 &  5.63 &   0.21 \\ 
\enddata
\tablecomments{Differential number counts measured for S-CANDELS measured in bins of width
0.5\,mag, expressed in terms of log$(N)$\,mag$^{-1}$\,deg$^{-2}$.  Counts given as 
``Total" are area-weighted means derived from all five S-CANDELS fields using the areas 
given in Table~\ref{tab:obslog}.  All uncertainties are 
1$\sigma$.  The errors given for individual fields reflect only $\sqrt N$ counting 
errors, but the uncertainties attributed to ``Total" counts also take field-field
variations into account.
}
\end{deluxetable*}

\subsection{The Integrated Background Light from IRAC Sources}

Space-based surveys such as S-CANDELS, hold the potential to
identify the source giving rise to the Cosmic Infrared Background (CIB).
That part of the CIB that arises in discrete sources can be robustly estimated
by identifying and photometering those sources and subsequently computing 
their contribution to the CIB {\sl in toto}.  Indeed, this was one of
the original motivations for both SEDS and S-CANDELS.  The outcome of the
SEDS measurement was that IRAC-detected sources account for only about 
half of the DIRBE CIB estimates.
More specifically, SEDS 3.6 and 4.5\,$\mu$m
sources account respectively for $5.6\pm1.0$ and $4.4\pm0.8$\,nW\,m$^{-2}$\,sr$^{-1}$ 
down to 26\,mag (see Ashby et al.\
for details of the measurement).  This is less than half the estimate from DIRBE 
(13.3$\pm2.8$\,nW\,m$^{-2}$sr$^{-1}$; Levenson \etal 2007),
although one should bear in mind that the DIRBE estimate depends on 
modeling and subtracting a large and inherently uncertain zodiacal 
foreground that is very bright compared to the CIB surface brightness.  

With its greater completeness, S-CANDELS confirms the initial SEDS 
measurements (Figures~\ref{fig:counts1} and \ref{fig:counts2}) of the
resolved fraction of the CIB.  However, the increase in the resolved CIB
from S-CANDELS is small: just
$0.08\pm0.03$ and $0.09\pm0.03$\,nW\,m$^{-2}$\,sr$^{-1}$ at 3.6 and 4.5\,$\mu$m,
respectively.  Thus even with the fourfold increase in overall integration time,
the marginal increase in resolved CIB light from S-CANDELS is much less
than the uncertainty in the original SEDS measurement.  The revised
estimate for the total contribution of resolved sources to the CIB in the
IRAC bands is $5.7\pm1.0$ and $4.5\pm0.8$\,nW\,m$^{-2}$\,sr$^{-1}$.

\section{S-CANDELS Catalogs}
\label{sec:catalogs}

The S-CANDELS IRAC catalogs are presented in Tables~\ref{tab:uds} through \ref{tab:egs}.  
In addition to the PSF-fitted magnitudes on which the source detection is based, 
they also list the IRAC positions and photometry at both 3.6 and 4.5\,$\mu$m 
in six apertures
of diameters $2\farcs4, 3\farcs6, 4\farcs8, 6\arcsec, 7\farcs2$, and 12\arcsec.
All the photometry has been aperture-corrected and adjusted to account for
the empirically determined biases given in Table~\ref{tab:bias}, i.e., all magnitudes
are expressed as total magnitudes.  The catalogs also provide 1$\sigma$ uncertainty 
estimates, but only for the $2\farcs4$-diameter aperture photometry, because time 
constraints made it impractical to simulate photometry for all the aperture diameters.
Users are encouraged to use the $2\farcs4$ aperture for photometry of faint sources.  
The other apertures are provided so that users can construct the 
curve of growth for large, extended sources.

Users should be aware of some limitations of the S-CANDELS catalogs, which
are described here.

\begin{itemize}

\item At the faintest levels ($[3.6]=[4.5]>26$) residual effects of cosmic rays
may, on average, lead sources to appear slightly brighter than they really 
are.  If real, this flux boosting is comparable in
magnitude to the cataloged uncertainties, but would not have been captured
in the simulations because the simulated sources were inserted into the final
mosaics, not the individual exposures.

\item Although a source must be detected in both IRAC bands in order to be
included in the S-CANDELS catalogs, those catalogs nonetheless contain 
some spurious sources (for example, where Airy rings of bright sources overlap).

\item Down to $[3.6]=[4.5]=26.5$ the S-CANDELS catalogs are limited more 
strongly by source confusion than by sensitivity, given the high source 
area density relative to the IRAC beam sizes at 3.6 and 4.5\,$\mu$m.  
It is therefore inevitable that some real IRAC sources lying well above 
the detection threshold are absent from the S-CANDELS catalogs 
(Section~\ref{ssec:extraction}).

\item Sources brighter than $[3.6]=[4.5]=15.4$ lie close to the IRAC saturation
limit, and their cataloged photometry is therefore suspect. 
\end{itemize}

\section{Applications of the S-CANDELS IRAC Data}
\label{sec:applications}

S-CANDELS was conceived and executed to aid in detecting and characterizing 
the faintest, most distant objects accessible in the 3.6 and 4.5\,$\mu$m 
bandpasses.  
In its survey area, S-CANDELS quadrupled the total integration time of its
predecessor SEDS from 12 to 50\,hours, but over a smaller area, just 0.16\,deg$^2$.  
In doing so,
S-CANDELS achieved significantly higher completeness for the sources most
likely to lie at high redshift, i.e., objects fainter than 24\,mag in the IRAC
bands.

The CANDELS collaboration has already
been combining S-CANDELS IRAC data with imaging from {\sl HST}/ACS and 
WFC3: in the ECDFS by Guo et al.\ (2013), in COSMOS by Nayyeri et al.\ in 
preparation, in the HDFN by Barro et al.\ in preparation, and in the EGS by
Stefanon et al.\ in preparation.  But at the same time the S-CANDELS data
have also been used in several studies that exploit their long-anticipated
utility for constraining
the properties of individual sources.  For example, Mortlock et al. (2015) 
combined S-CANDELS with CANDELS data to estimate the galaxy stellar mass
function for galaxies in the redshift range $0.3<z<3$.  
Duncan et al.\ (2014) and Grazian et al.\ (2015) carried out related
analyses but at more distant redshifts, $4<z<7$ and $3.5<z<7.5$, respectively.  
These and similar efforts exploit the special power of IRAC
photometry to elucidate the stellar masses of distant objects.
Smit et al.\ (2015) exploited IRAC S-CANDELS photometry to identify promising 
galaxy candidates 
in a high but narrow redshift range, $6.6<z<6.9$, interpreting
their rare and very blue IRAC colors as the effect of strong nebular emission.
The S-CANDELS IRAC photometry is also useful for constraining photometric
redshifts because it extends the CANDELS
coverage into the rest-frame near-infrared wavelengths for distant galaxies 
(Nayyeri et al.\ 2014) or even rest-visible wavelengths for galaxies at
extreme redshifts (Ouchi et al. 2013; Hsu et al.\ 2014; Finkelstein et al.\ 2014; 2015).
These achievements hint at a potentially rich legacy for S-CANDELS, 
but it is very likely that other projects not yet even imagined will also
make use of these data in the years to come.

\acknowledgments

The authors are grateful to E. Diolaiti for helpful advice on the 
optimal use of {\tt StarFinder}.
This work is based on observations made with the {\it Spitzer Space
Telescope}, which is operated by the Jet Propulsion Laboratory,
California Institute of Technology under contract with the National 
Aeronautics and Space Administration (NASA).  Support for this 
work was provided by NASA through contract numbers 1439337 and 1439801 
issued by JPL/Caltech, and by {\sl HST} grant GO-12060.05-A.  
JSD acknowledges the support of the European
Research Council via an Advanced Grant, and the support of the Royal 
Society via a Wolfson Research Merit Award.
IRAF is distributed by the National Optical Astronomy 
Observatory, which is operated by the Association of Universities for 
Research in Astronomy (AURA) under cooperative agreement with the 
National Science Foundation.  

Facilities:  \facility{{\it Spitzer Space Telescope} (IRAC)}

{}
\clearpage

\begin{deluxetable}{ccccccccccclc}
\tabletypesize{\scriptsize}
\tablecaption{Full-Depth Source Catalog for the S-CANDELS UDS Field\label{tab:uds}}
\tablehead{
\colhead{Object} &
\colhead{RA,Dec} &
\colhead{3.6\,$\mu$m AB Magnitudes\tablenotemark{a}} &
\colhead{3.6\,$\mu$m Unc.\tablenotemark{b}} &
\colhead{3.6\,$\mu$m Bias\tablenotemark{c}} &
\colhead{3.6\,$\mu$m Coverage\tablenotemark{d}} \\ 
&
\colhead{(J2000)}&
\colhead{4.5\,$\mu$m AB Magnitudes} &
\colhead{4.5\,$\mu$m Unc.} &
\colhead{4.5\,$\mu$m Bias\tablenotemark{e}} &
\colhead{4.5\,$\mu$m Coverage\tablenotemark{f}} \\
}
\startdata
SCANDELS J021756.87-050757.4 &  34.48698,-5.13260 &  14.03  14.02  13.85  13.78  13.74  13.72  13.68 &   0.03 &   0.00 &   197 \\ 
                             &                    &  14.12  14.12  14.05  13.98  13.94  13.94  13.92 &   0.03 &   0.00 &   198 \\ 
SCANDELS J021725.24-051804.6 &  34.35517,-5.30129 &  14.25  14.24  14.16  14.11  14.09  14.08  14.06 &   0.03 &   0.00 &   203 \\ 
                             &                    &  14.52  14.52  14.49  14.44  14.41  14.41  14.41 &   0.03 &   0.00 &   384 \\ 
SCANDELS J021657.23-050801.5 &  34.23847,-5.13375 &  14.29  14.29  14.20  14.16  14.14  14.13  14.11 &   0.03 &   0.00 &   253 \\ 
                             &                    &  14.53  14.57  14.56  14.55  14.55  14.57  14.60 &   0.03 &   0.00 &   335 \\ 
SCANDELS J021654.28-051817.3 &  34.22616,-5.30482 &  14.35  14.34  14.31  14.29  14.28  14.28  14.26 &   0.03 &   0.00 &   236 \\ 
                             &                    &  14.58  14.58  14.61  14.60  14.59  14.59  14.60 &   0.03 &   0.00 &   297 \\ 
SCANDELS J021803.06-051628.9 &  34.51275,-5.27470 &  14.40  14.17  13.99  13.84  13.65  13.50  13.00 &   0.03 &   0.00 &   193 \\ 
                             &                    &  13.90  13.78  13.66  13.60  13.53  13.45  13.23 &   0.03 &   0.00 &   179 \\ 
SCANDELS J021823.63-051923.9 &  34.59845,-5.32332 &  14.53  14.53  14.53  14.53  14.53  14.53  14.53 &   0.03 &   0.00 &   266 \\ 
                             &                    &  14.86  14.86  14.90  14.90  14.89  14.89  14.90 &   0.03 &   0.00 &   328 \\ 
SCANDELS J021724.98-051320.1 &  34.35407,-5.22225 &  14.58  14.57  14.52  14.49  14.47  14.46  14.45 &   0.03 &   0.00 &   323 \\ 
                             &                    &  14.78  14.78  14.81  14.79  14.77  14.76  14.77 &   0.03 &   0.00 &   516 \\ 
SCANDELS J021649.03-051556.9 &  34.20429,-5.26580 &  14.71  14.71  14.75  14.77  14.78  14.78  14.79 &   0.03 &   0.00 &   331 \\ 
                             &                    &  15.08  15.08  15.15  15.16  15.16  15.16  15.21 &   0.03 &   0.00 &   408 \\ 
SCANDELS J021721.60-050935.6 &  34.34002,-5.15988 &  14.80  14.80  14.81  14.81  14.82  14.82  14.82 &   0.03 &   0.00 &  1030 \\ 
                             &                    &  15.04  15.04  15.12  15.13  15.13  15.14  15.16 &   0.03 &   0.00 &  1918 \\ 
\enddata
\tablecomments{The S-CANDELS catalog of sources in the UDS field selected at both
3.6 and 4.5\,\micron\ as described in the text.  The sources are listed in 
magnitude order.
This table is available in its entirety in a machine-readable form in the online
journal.  A portion is shown here for guidance regarding its form and content.
}
\tablenotetext{a}{The PSF-fitted magnitude is given first, and the magnitudes
given after are measured in apertures of 2\farcs4, 3\farcs6, 4\farcs8, 6\farcs0,
7\farcs2, and 12\farcs0 diameter, corrected to total.}
\tablenotetext{b}{Uncertainties given are 1$\sigma$, and apply to the 2\farcs4
diameter aperture magnitudes.}
\tablenotetext{c}{3.6\,$\mu$m photometric biases already applied to the aperture photometry.}
\tablenotetext{d}{Depth of coverage expressed in units of 100\,s
IRAC frames that observed the source.}
\end{deluxetable}

\begin{deluxetable}{ccccccccccccclc}
\tabletypesize{\scriptsize}
\tablecaption{Full-Depth Source Catalog for the S-CANDELS ECDFS Field\label{tab:ecdfs}}
\tablehead{
\colhead{Object} &
\colhead{RA,Dec} &
\colhead{3.6\,$\mu$m AB Magnitudes\tablenotemark{a}} &
\colhead{3.6\,$\mu$m Unc.\tablenotemark{b}} &
\colhead{3.6\,$\mu$m Bias\tablenotemark{c}} &
\colhead{3.6\,$\mu$m Coverage\tablenotemark{d}} \\ 
&
\colhead{(J2000)}&
\colhead{4.5\,$\mu$m AB Magnitudes} &
\colhead{4.5\,$\mu$m Unc.} &
\colhead{4.5\,$\mu$m Bias\tablenotemark{e}} &
\colhead{4.5\,$\mu$m Coverage\tablenotemark{f}} \\
}
\startdata
SCANDELS J033314.06-273424.8 &  53.30857,-27.57356 &  11.22   11.20   11.12   11.11   11.11   11.11   11.12 &  0.03  & 0.00 &   420 \\ 
                             &                     &  11.87   11.82   11.72   11.70   11.68   11.67   11.67 &  0.03  & 0.00 &   381 \\ 
SCANDELS J033222.57-275805.6 &  53.09403,-27.96822 &  12.19   12.16   12.11   12.10   12.10   12.10   12.10 &  0.03  & 0.00 &   794 \\ 
                             &                     &  12.84   12.79   12.71   12.68   12.66   12.65   12.65 &  0.03  & 0.00 &   729 \\ 
SCANDELS J033242.07-275702.2 &  53.17528,-27.95062 &  12.78   12.74   12.70   12.69   12.69   12.69   12.70 &  0.03  & 0.00 &   749 \\ 
                             &                     &  13.47   13.44   13.35   13.30   13.28   13.26   13.25 &  0.03  & 0.00 &   706 \\ 
SCANDELS J033316.93-275338.7 &  53.32054,-27.89409 &  13.76   13.74   13.69   13.66   13.65   13.65   13.66 &  0.03  & 0.00 &  1586 \\ 
                             &                     &  14.46   14.43   14.36   14.33   14.31   14.31   14.30 &  0.03  & 0.00 &   505 \\ 
SCANDELS J033314.46-275428.0 &  53.31027,-27.90777 &  14.55   14.54   14.52   14.51   14.52   14.52   14.52 &  0.03  & 0.00 &  1580 \\ 
                             &                     &  14.85   14.86   14.90   14.91   14.91   14.92   14.92 &  0.03  & 0.00 &   468 \\ 
SCANDELS J033219.13-273933.6 &  53.07972,-27.65933 &  14.62   14.59   14.53   14.51   14.50   14.50   14.49 &  0.03  & 0.00 &   503 \\ 
                             &                     &  14.75   14.76   14.78   14.80   14.81   14.82   14.83 &  0.03  & 0.00 &   503 \\ 
SCANDELS J033318.60-274218.5 &  53.32752,-27.70513 &  14.66   14.65   14.61   14.60   14.60   14.60   14.60 &  0.03  & 0.00 &   457 \\ 
                             &                     &  14.89   14.89   14.92   14.93   14.93   14.94   14.94 &  0.03  & 0.00 &   444 \\ 
SCANDELS J033312.35-274232.8 &  53.30144,-27.70911 &  14.74   14.72   14.69   14.68   14.68   14.67   14.67 &  0.03  & 0.00 &   484 \\ 
                             &                     &  15.01   15.03   15.06   15.08   15.10   15.12   15.14 &  0.03  & 0.00 &   475 \\ 
SCANDELS J033159.82-274917.0 &  52.99924,-27.82140 &  14.79   14.78   14.76   14.75   14.75   14.76   14.76 &  0.03  & 0.00 &   461 \\ 
                             &                     &  15.05   15.06   15.07   15.08   15.09   15.09   15.10 &  0.03  & 0.00 &   503 \\ 
\enddata
\tablecomments{The S-CANDELS catalog of sources in the ECDFS field selected at both
3.6 and 4.5\,\micron\ as described in the text.  The sources are listed in 
magnitude order.
This table is available in its entirety in a machine-readable form in the online
journal.  A portion is shown here for guidance regarding its form and content.
}
\tablenotetext{a}{The PSF-fitted magnitude is given first, and the magnitudes
given after are measured in apertures of 2\farcs4, 3\farcs6, 4\farcs8, 6\farcs0,
7\farcs2, and 12\farcs0 diameter, corrected to total.}
\tablenotetext{b}{Uncertainties given are 1$\sigma$, and apply to the 2\farcs4
diameter aperture magnitudes.}
\tablenotetext{c}{3.6\,$\mu$m photometric biases already applied to the aperture photometry.}
\tablenotetext{d}{Depth of coverage expressed in units of 100\,s
IRAC frames that observed the source.}
\end{deluxetable}

\begin{deluxetable}{ccccccccccccclc}
\tabletypesize{\scriptsize}
\tablecaption{Full-Depth Source Catalog for the S-CANDELS COSMOS Field\label{tab:cosmos}}
\tablehead{
\colhead{Object} &
\colhead{RA,Dec} &
\colhead{3.6\,$\mu$m AB Magnitudes\tablenotemark{a}} &
\colhead{3.6\,$\mu$m Unc.\tablenotemark{b}} &
\colhead{3.6\,$\mu$m Bias\tablenotemark{c}} &
\colhead{3.6\,$\mu$m Coverage\tablenotemark{d}} \\ 
&
\colhead{(J2000)}&
\colhead{4.5\,$\mu$m AB Magnitudes} &
\colhead{4.5\,$\mu$m Unc.} &
\colhead{4.5\,$\mu$m Bias\tablenotemark{e}} &
\colhead{4.5\,$\mu$m Coverage\tablenotemark{f}} \\
}
\startdata
SCANDELS J100009.66+022349.0 & 150.04023,2.39693  &  11.10   11.06   11.00   10.96   10.95   10.95   10.95 &  0.03 &  0.00 &  1267 \\ 
                             &                    &  11.67   11.62   11.53   11.51   11.50   11.49   11.48 &  0.03 &  0.00 &   555 \\ 
SCANDELS J100002.32+023259.2 & 150.00969,2.54979  &  12.56   12.55   12.54   12.53   12.51   12.51   12.51 &  0.03 &  0.00 &     6 \\ 
                             &                    &  13.90   13.82   13.64   13.52   13.46   13.44   13.40 &  0.03 &  0.00 &    10 \\ 
SCANDELS J100032.57+020825.6 & 150.13569,2.14045  &  12.69   12.66   12.61   12.59   12.57   12.56   12.56 &  0.03 &  0.00 &   357 \\ 
                             &                    &  13.27   13.24   13.18   13.13   13.11   13.10   13.09 &  0.03 &  0.00 &   295 \\ 
SCANDELS J100036.89+022357.5 & 150.15371,2.39930  &  13.60   13.58   13.53   13.50   13.49   13.48   13.48 &  0.03 &  0.00 &  2091 \\ 
                             &                    &  14.43   14.40   14.31   14.26   14.23   14.22   14.20 &  0.03 &  0.00 &  1104 \\ 
SCANDELS J100027.69+022752.3 & 150.11539,2.46452  &  13.74   13.72   13.66   13.63   13.62   13.61   13.61 &  0.03 &  0.00 &  1138 \\ 
                             &                    &  14.57   14.54   14.42   14.37   14.34   14.34   14.32 &  0.03 &  0.00 &   616 \\ 
SCANDELS J100104.31+023015.9 & 150.26796,2.50441  &  13.90   13.78   13.57   13.43   13.34   13.30   13.24 &  0.03 &  0.00 &    12 \\ 
                             &                    &  13.85   13.78   13.63   13.53   13.47   13.46   13.42 &  0.03 &  0.00 &     6 \\ 
SCANDELS J100017.19+022554.9 & 150.07163,2.43191  &  14.00   13.99   13.93   13.88   13.87   13.86   13.85 &  0.03 &  0.00 &   804 \\ 
                             &                    &  14.54   14.51   14.44   14.40   14.38   14.37   14.35 &  0.03 &  0.00 &   807 \\ 
SCANDELS J095954.72+021706.6 & 149.97801,2.28518  &  14.10   14.07   13.95   13.89   13.85   13.84   13.82 &  0.03 &  0.00 &     9 \\ 
                             &                    &  14.30   14.28   14.22   14.19   14.17   14.16   14.15 &  0.03 &  0.00 &     7 \\ 
SCANDELS J100045.10+021636.9 & 150.18790,2.27693  &  14.15   14.12   14.05   14.00   13.97   13.96   13.95 &  0.03 &  0.00 &   662 \\ 
                             &                    &  14.58   14.55   14.50   14.47   14.45   14.44   14.43 &  0.03 &  0.00 &   719 \\ 
\enddata
\tablecomments{The S-CANDELS catalog of sources in the COSMOS field selected at both
3.6 and 4.5\,\micron\ as described in the text.  The sources are listed in 
magnitude order.
This table is available in its entirety in a machine-readable form in the online
journal.  A portion is shown here for guidance regarding its form and content.
}
\tablenotetext{a}{The PSF-fitted magnitude is given first, and the magnitudes
given after are measured in apertures of 2\farcs4, 3\farcs6, 4\farcs8, 6\farcs0,
7\farcs2, and 12\farcs0 diameter, corrected to total.}
\tablenotetext{b}{Uncertainties given are 1$\sigma$, and apply to the 2\farcs4
diameter aperture magnitudes.}
\tablenotetext{c}{3.6\,$\mu$m photometric biases already applied to the aperture photometry.}
\tablenotetext{d}{Depth of coverage expressed in units of 100\,s
IRAC frames that observed the source.}
\end{deluxetable}

\begin{deluxetable}{ccccccccccccclc}
\tabletypesize{\scriptsize}
\tablecaption{Full-Depth Source Catalog for the S-CANDELS HDFN Field\label{tab:hdfn}}
\tablehead{
\colhead{Object} &
\colhead{RA,Dec} &
\colhead{3.6\,$\mu$m AB Magnitudes\tablenotemark{a}} &
\colhead{3.6\,$\mu$m Unc.\tablenotemark{b}} &
\colhead{3.6\,$\mu$m Bias\tablenotemark{c}} &
\colhead{3.6\,$\mu$m Coverage\tablenotemark{d}} \\ 
&
\colhead{(J2000)}&
\colhead{4.5\,$\mu$m AB Magnitudes} &
\colhead{4.5\,$\mu$m Unc.} &
\colhead{4.5\,$\mu$m Bias\tablenotemark{e}} &
\colhead{4.5\,$\mu$m Coverage\tablenotemark{f}} \\
}
\startdata
SCANDELS J123737.90+621630.6 & 189.40794,62.27517  &  12.97   12.96   12.97   12.97   12.97   12.97   12.99 &  0.03  & 0.00 &  2031 \\ 
                             &                     &  13.85   13.82   13.75   13.69   13.66   13.65   13.64 &  0.03  & 0.00 &   812 \\ 
SCANDELS J123653.00+620727.1 & 189.22084,62.12419  &  13.12   13.12   13.12   13.11   13.10   13.10   13.10 &  0.03  & 0.00 &   862 \\ 
                             &                     &  14.06   14.03   13.91   13.84   13.80   13.79   13.77 &  0.03  & 0.00 &   476 \\ 
SCANDELS J123625.05+622115.8 & 189.10438,62.35439  &  14.28   14.28   14.15   14.08   14.04   14.03   14.00 &  0.03  & 0.00 &    70 \\ 
                             &                     &  14.56   14.52   14.44   14.39   14.36   14.35   14.33 &  0.03  & 0.00 &   211 \\ 
SCANDELS J123640.15+621941.4 & 189.16728,62.32817  &  14.62   14.61   14.57   14.54   14.52   14.51   14.51 &  0.03  & 0.00 &  1047 \\ 
                             &                     &  15.08   15.04   14.98   14.95   14.93   14.92   14.92 &  0.03  & 0.00 &  1341 \\ 
SCANDELS J123554.73+622201.8 & 188.97804,62.36716  &  14.80   14.18   13.95   13.74   13.47   13.27   12.70 &  0.03  & 0.00 &   240 \\ 
                             &                     &  14.50   14.16   13.95   13.81   13.61   13.42   12.97 &  0.03  & 0.00 &    21 \\ 
SCANDELS J123536.12+621647.2 & 188.90052,62.27976  &  14.80   14.81   14.81   14.81   14.81   14.81   14.82 &  0.03  & 0.00 &   298 \\ 
                             &                     &  15.12   15.13   15.15   15.16   15.17   15.18   15.19 &  0.03  & 0.00 &   504 \\ 
SCANDELS J123743.03+621900.9 & 189.42929,62.31692  &  14.86   14.84   14.82   14.81   14.80   14.80   14.80 &  0.03  & 0.00 &  1357 \\ 
                             &                     &  15.13   15.18   15.21   15.24   15.28   15.30   15.37 &  0.03  & 0.00 &  1795 \\ 
SCANDELS J123546.52+620749.2 & 188.94385,62.13033  &  14.90   14.91   14.94   14.96   14.97   14.97   14.97 &  0.03  & 0.00 &   402 \\ 
                             &                     &  15.40   15.40   15.42   15.43   15.44   15.44   15.44 &  0.03  & 0.00 &   508 \\ 
SCANDELS J123633.72+620807.3 & 189.14049,62.13535  &  14.96   14.96   14.97   14.98   14.98   14.98   14.98 &  0.03  & 0.00 &  1219 \\ 
                             &                     &  15.32   15.35   15.39   15.41   15.43   15.43   15.45 &  0.03  & 0.00 &  1212 \\ 
\enddata
\tablecomments{The S-CANDELS catalog of sources in the HDFN field selected at both
3.6 and 4.5\,\micron\ as described in the text.  The sources are listed in 
magnitude order.
This table is available in its entirety in a machine-readable form in the online
journal.  A portion is shown here for guidance regarding its form and content.
}
\tablenotetext{a}{The PSF-fitted magnitude is given first, and the magnitudes
given after are measured in apertures of 2\farcs4, 3\farcs6, 4\farcs8, 6\farcs0,
7\farcs2, and 12\farcs0 diameter, corrected to total.}
\tablenotetext{b}{Uncertainties given are 1$\sigma$, and apply to the 2\farcs4
diameter aperture magnitudes.}
\tablenotetext{c}{3.6\,$\mu$m photometric biases already applied to the aperture photometry.}
\tablenotetext{d}{Depth of coverage expressed in units of 100\,s
IRAC frames that observed the source.}
\end{deluxetable}

\begin{deluxetable}{ccccccccccclc}
\tabletypesize{\scriptsize}
\tablecaption{Full-Depth Source Catalog for the S-CANDELS EGS Field\label{tab:egs}}
\tablehead{
\colhead{Object} &
\colhead{RA,Dec} &
\colhead{3.6\,$\mu$m AB Magnitudes\tablenotemark{a}} &
\colhead{3.6\,$\mu$m Unc.\tablenotemark{b}} &
\colhead{3.6\,$\mu$m Bias\tablenotemark{c}} &
\colhead{3.6\,$\mu$m Coverage\tablenotemark{d}} \\ 
&
\colhead{(J2000)}&
\colhead{4.5\,$\mu$m AB Magnitudes} &
\colhead{4.5\,$\mu$m Unc.} &
\colhead{4.5\,$\mu$m Bias\tablenotemark{e}} &
\colhead{4.5\,$\mu$m Coverage\tablenotemark{f}} \\
}
\startdata
SCANDELS J141904.58+524811.5 & 214.76910,52.80319 &  13.45   13.45   13.41   13.40   13.39   13.39   13.40  & 0.03  & 0.00 &  1567 \\ 
                             &                    &  14.46   14.42   14.32   14.26   14.23   14.22   14.21  & 0.03  & 0.00 &   743 \\ 
SCANDELS J141846.53+524522.2 & 214.69387,52.75616 &  14.73   14.74   14.70   14.68   14.67   14.66   14.66  & 0.03  & 0.00 &   640 \\ 
                             &                    &  14.84   14.85   14.89   14.91   14.92   14.92   14.93  & 0.03  & 0.00 &  1141 \\ 
SCANDELS J142101.89+530316.4 & 215.25787,53.05455 &  14.76   14.76   14.73   14.72   14.72   14.71   14.71  & 0.03  & 0.00 &   679 \\ 
                             &                    &  14.96   14.98   15.02   15.04   15.06   15.06   15.07  & 0.03  & 0.00 &  1012 \\ 
SCANDELS J142009.19+525508.9 & 215.03830,52.91914 &  14.91   14.95   14.99   15.02   15.04   15.06   15.12  & 0.03  & 0.00 &  2434 \\ 
                             &                    &  15.37   15.39   15.44   15.46   15.47   15.47   15.49  & 0.03  & 0.00 &  2487 \\ 
SCANDELS J142046.77+530329.7 & 215.19486,53.05825 &  14.97   14.97   15.01   15.03   15.03   15.04   15.04  & 0.03  & 0.00 &  1737 \\ 
                             &                    &  15.43   15.45   15.49   15.52   15.53   15.54   15.56  & 0.03  & 0.00 &  1840 \\ 
SCANDELS J142002.64+530118.2 & 215.01100,53.02171 &  15.15   15.15   15.19   15.20   15.21   15.21   15.22  & 0.03  & 0.00 &   524 \\ 
                             &                    &  15.59   15.60   15.63   15.65   15.66   15.67   15.67  & 0.03  & 0.00 &   562 \\ 
SCANDELS J142053.38+530015.0 & 215.22240,53.00418 &  15.16   15.16   15.21   15.22   15.23   15.23   15.24  & 0.03  & 0.00 &   603 \\ 
                             &                    &  15.63   15.64   15.68   15.69   15.70   15.70   15.71  & 0.03  & 0.00 &   605 \\ 
SCANDELS J141907.48+524630.1 & 214.78116,52.77502 &  15.18   15.19   15.23   15.24   15.25   15.25   15.26  & 0.03  & 0.00 &  1548 \\ 
                             &                    &  15.70   15.70   15.74   15.75   15.76   15.76   15.75  & 0.03  & 0.00 &  1580 \\ 
SCANDELS J141941.02+525108.5 & 214.92090,52.85235 &  15.33   15.36   15.40   15.42   15.44   15.45   15.49  & 0.03  & 0.00 &  1773 \\ 
                             &                    &  15.84   15.86   15.89   15.91   15.92   15.93   15.94  & 0.03  & 0.00 &  1827 \\ 
\enddata
\tablecomments{The S-CANDELS catalog of sources in the EGS field selected at both
3.6 and 4.5\,\micron\ as described in the text.  The sources are listed in 
magnitude order.
This table is available in its entirety in a machine-readable form in the online
journal.  A portion is shown here for guidance regarding its form and content.
}
\tablenotetext{a}{The PSF-fitted magnitude is given first, and the magnitudes
given after are measured in apertures of 2\farcs4, 3\farcs6, 4\farcs8, 6\farcs0,
7\farcs2, and 12\farcs0 diameter, corrected to total.}
\tablenotetext{b}{Uncertainties given are 1$\sigma$, and apply to the 2\farcs4
diameter aperture magnitudes.}
\tablenotetext{c}{3.6\,$\mu$m photometric biases already applied to the aperture photometry.}
\tablenotetext{d}{Depth of coverage expressed in units of 100\,s
IRAC frames that observed the source.}
\end{deluxetable}


\begin{thebibliography}{}

\bibitem[Ashby et al.(2009)]{2009ash}
Ashby, M., et al.\ 2009, \apj, 701, 428

\bibitem[Ashby et al.(2013)]{2013ash}
Ashby, M., et al.\ 2013, \apj, 769, 80 

\bibitem[Ashby et al.(2013b)]{2013bash}
Ashby, M., et al.\ 2013b, \apjs, 209, 228

\bibitem[Arendt et al.(1998)]{1996are}
Arendt, R G., et al.\ 1998, \apj, 508, 74

\bibitem[Barmby \etal (2008)]{2008bar} 
Barmby, P., et al.\ 2008, \apjs, 177, 431

\bibitem[Barro et al.(2011)]{2011bar}
Barro, G., et al.\ 2011, \apjs, 193, 30

\bibitem[Bielby al.(2012)]{2012A&A...545A..23B} 
Bielby, R., Hudelot, P., McCracken, H.~J., et al.\ 2012, \aap, 545, A23 

\bibitem[Caputi et al.(2011)]{2011MNRAS.413..162C} 
Caputi, K., et al.\ 2011, \mnras, 413, 162

\bibitem[Castellano et al.(2010)]{2010A&A...511A..20C} 
Castellano, M., Fontana, A., Boutsia, K., et al.\ 2010, \aap, 511, A20 

\bibitem[Cohen et al.(1993)]{1993coh}
Cohen, M., et al.\ 1993, \aj, 105, 1860

\bibitem[Cohen et al.(1994)]{1994coh}
Cohen, M., et al.\ 1994, \aj, 107, 582

\bibitem[Cohen et al.(1995)]{1995coh}
Cohen, M., et al.\ 1995, \apj, 444, 874

\bibitem[Damen et al.(2011)]{2011dam}
Damen, M., et al.\ 2011, \apj, 727, 1

\bibitem[Davis et al.(2007)]{2007ApJ...660L...1D} 
Davis, M., et al.\ 2007, \apjl, 660, L1 

\bibitem[Diolaiti et al.\ (2000)]{2000SPIE.4007..879D}
Diolaiti, E., \etal 2000, \procspie, 4007, 879

\bibitem[Duncan et al.\ (2014)]{2014dun}
Duncan, K., Conselice, C.~J., Mortlock, A., et al.\ 2014, \mnras, 444, 2960

\bibitem[Fazio et al.\ (2004a)]{2004ApJS..154...10F} 
Fazio, G.~G., et al.\ 2004a, \apjs, 154, 10

\bibitem[Fazio et al.\ (2004b)]{2004ApJS..154...39F} 
Fazio, G.~G., et al.\ 2004b, \apjs, 154, 39 

\bibitem[Fang et al.\ (2004)]{2004fan}
Fang, F., et al.\ 2004, \apjs, 154, 35

\bibitem[Finkelstein et al.\ (2015)]{2015fin}
Finkelstein, S.~L., Ryan, R.~E.~Jr., Papovich, C., et al.\ 2014, arXiv:1410.5439

\bibitem[Finkelstein et al.\ (2015)]{2013fin}
Finkelstein, S.~L., Papovich, C., Dickinson, M., et al.\ 2013, {\sl Nature}, 502, 524

\bibitem[Franceschini et al.\ (2006)]{Franceschini2006}
Franceschini, A., Rodighiero, G., Cassata, P., et al.\ 2006, \aap, 397

\bibitem[Galametz et al.(2013)]{2013gal}
Galametz, Z., Grazian, A., Fontana, A., et al.\ 2013, \apjs, 206, 10

\bibitem[Giavalisco et al.(2004)]{2004gia}
Giavalisco, M., Ferguson, H.~C., Koekemoer, A.~M., et al.\ 2004, \apjl, 600, L93

\bibitem[Grazian et al.(2015)]{2015gra}
Grazian, A., Fontana, A., Santini, P., et al.\ 2015, \aap accepted, arXiv:1412.0532

\bibitem[Grogin et al.(2011)]{2011ApJS..197...35G} Grogin, N.~A., et al.\ 
2011, \apjs, 197, 35  

\bibitem[Guo et al.(2013)]{2013guo}
Guo, Y., Ferguson, H., Giavalisco, M., et al.\ 2013, \apjs, 207, 24

\bibitem[Hathi et al.(2012)]{2012ApJ...757...43H} 
Hathi, N.~P., Mobasher, B., Capak, P., Wang, W.-H., \& Ferguson, H.~C.\ 2012, \apj, 757, 43 

\bibitem[Helgason \etal (2012)]{Helgason2012}
Helgason, K., Ricotti, M., \& Kashlinsky, A.\ 2012, \apj, 752, 113

\bibitem[Hsu et al. 2014]{2014hsu}
Hsu, L.-T., Salvato, M., Nandra, K., et al.\ 2014, \apj, 796, 60.

\bibitem[Koekemoer et al.(2007)]{2007koe}
Koekemoer, A.~M., Aussel, H., Calzetti, D., et al.\ 2007, \apjs, 172, 96  

\bibitem[Koekemoer et al.(2011)]{2011ApJS..197...36K} Koekemoer, A.~M., et 
al.\ 2011, \apjs, 197, 36  

\bibitem[Labb\'e et al.(2013)]{2012lab}
Labb\'e, I., Oesch, P.~A., Bouwens, R.~J., et al.\ 2013, \apj, 777, 19

\bibitem[Lawrence \etal (2007)]{2007law} 
Lawrence, A., Warren, S.~J., Almaini, O., et al.\ 2007, \mnras, 379, 1599

\bibitem[Levenson \etal (2007)]{2007lev} 
Levenson, L.~R., Wright, E.~L., \& Johnson, B.~D., 2007, \apj, 666, 34

\bibitem[Lin et al.(2012)]{2010lin} 
Lin, L., Dickinson, M., Jian, H.-Y., et al. 2012, \apj, 756, 71

\bibitem[Lonsdale et al.(2003)]{2003PASP..115...897L} 
Lonsdale, C.~J., et al.\ 2003, \pasp, 115, 897 

\bibitem[Lonsdale et al.(2004)]{2004ApJS..154...54L} 
Lonsdale, C.~J., et al.\ 2004, \apjs, 154, 54 

\bibitem[Lotz et al.(2008)]{2008lot}
Lotz, J.~M., Davis, M., Faber, S.~M., et al.\ 2008, \apj, 672, 177

\bibitem[Magdis et al.\ (2008)]{2008mag}
Magdis, G.~E., Rigopoulou, D., Huang, J.-S., et al.\ 2008, \mnras, 386, 11

\bibitem[Mauduit et al.\ (2012)]{2012mau}
Mauduit, J.-C., Lacy, M., Farrah, D., et al.\ 2012, \pasp, 124, 1135

\bibitem[McCracken et al.(2012)]{2012mcc}
McCracken, H.~J., Milvang-Jensen, B., Dunlop, J., et al.\ 2012, \aap, 544, A156 

\bibitem[Mortlock et al.\ (2015)]{2015mor}
Mortlock, A., Conselice, C.~J., Hartley, W.~G., et al.\ 2015, \mnras, 447, 2

\bibitem[Nayyeri et al.(2014)]{2014nay}
Nayyeri, H., Mobasher, B., Hemmati, S., et al.\ 2014, \apj, 794, 68

\bibitem[Ouchi et al.(2001)]{2001ouc}
Ouchi, M., Shimasaku, K., Okamura, S., et al.\ 2001, \apjl, 558, L83

\bibitem[Ouchi et al.(2013)]{2013ouc}
Ouchi, M., Ellis, R., Ono, Y., et al.\ 2013, \apj, 778, 102 

\bibitem[Rix et al.(2004)]{2004ApJS..152..163R} 
Rix, H.-W., et al.\ 2004, \apjs, 152, 163 

\bibitem[Sanders et al.(2007)]{2007ApJS..172...86S} Sanders, D.~B., 
Salvato, M., Aussel, H., et al.\ 2007, \apjs, 172, 86 

\bibitem[Schuster et al.(2006)]{2006SPIE.6270E..65S} 
Schuster, M.~T., Marengo, M., \& Patten, B.~M.\ 2006, \procspie, 6270, 65

\bibitem[Scoville et al.(2007a)]{2007sco}
Scoville, N., Aussel H., Brusa, M., et al.\ 2007a, \apjs, 172, 1

\bibitem[Scoville et al.(2007b)]{2007ApJS..172...38S} 
Scoville, N., Abraham, R.~G., Aussel, H., et al.\ 2007b, \apjs, 172, 38 

\bibitem[Skrutskie et al.(2006)]{2006AJ....131.1163S} 
Skrutskie, M.~F., et al.\ 2006, \aj, 131, 1163

\bibitem[Smit et al.(2015)]{2015smi}
Smit, R., Bouwens, R.~J., Franx, M., et al.\ 2014, in press, arXiv:1412.0663

\bibitem[Sorba \& Sawicki(2010)]{Sorba2010}
Sorba, R., \& Sawicki, M.\ 2010, \apj, 721, 1056

\bibitem[Wainscoat et al.\ (1992)]{1992wai}
Wainscoat, R., et al.\ 1992, \apjs, 88, 529

\bibitem[Wang et al.(2010)]{2010wan}
Wang, W.-H., Cowie, L.~L., Barger, A.~J., Keenan, R.~C., \& Ting, H.-C. 2010, \apjs, 187, 251

\bibitem[Werner etal.(2004)]{2004wer}
Werner \etal 2004, \apjs, 154, 1

\end{thebibliography}
\end{document}